\documentclass[12pt,a4paper]{article}
\usepackage{jcappub}
\usepackage{amsmath}
\usepackage{amssymb}
\usepackage{epsfig}
\usepackage{latexsym,amsmath,amssymb}
\usepackage{graphicx}
\usepackage{slashed}
\usepackage{bm}
\usepackage{epsfig}
\usepackage{dcolumn}
\usepackage{color}

\def\beq{\begin{equation}}
\def\eeq{\end{equation}}
\def\ber{\begin{eqnarray}}
\def\eer{\end{eqnarray}}
\def\benu{\begin{enumerate}}
\def\eenu{\end{enumerate}}
\def\nn{\nonumber}
\def\l{\left}
\def\r{\right}
\def\d{{\rm d}}

\def\pa{\partial}
\def\f{\frac}
\def\mpl{M_{p}}

\def \lleq {\lower0.9ex\hbox{ $\buildrel < \over \sim$} ~}
\def \ggeq {\lower0.9ex\hbox{ $\buildrel > \over \sim$} ~}

\def\prl{{Phys.\@ Rev.\@ Lett.\ }}
\def\prd{{Phys.\@ Rev.\@ D\ }}

\def\plb {{Phys.\@ Lett.\@ B\ }}

\def\etal{{\it et al.}}
\def\ie {{\it ie}}

\title{Refining inflation using non-canonical scalars}
\author[a]{Sanil Unnikrishnan,}
\author[a]{Varun Sahni}
\author[b]{and Aleksey Toporensky}

\affiliation[a]{Inter-University Centre for Astronomy and Astrophysics,
Post Bag 4, Ganeshkhind, Pune 411~007, India}

\affiliation[b]{Sternberg Astronomical Institute, Moscow State University,
Universitetsky Prospekt, 13, Moscow 119992, Russia}

\emailAdd{sanil@iucaa.ernet.in}
\emailAdd{varun@iucaa.ernet.in}
\emailAdd{atopor@rambler.ru}

\date{\today}

\abstract{This paper revisits the Inflationary scenario within the framework of scalar field models possessing a non-canonical kinetic term. We obtain closed form solutions for all essential quantities associated with chaotic inflation including slow roll parameters, scalar and tensor power spectra, spectral indices, the tensor-to-scalar ratio, etc. We also examine the Hamilton-Jacobi equation and demonstrate the existence of an inflationary
attractor. Our results highlight the fact that non-canonical scalars can significantly improve the viability of inflationary models. They accomplish this by decreasing the tensor-to-scalar ratio while simultaneously increasing the value of the scalar spectral index, thereby redeeming models which are incompatible with the cosmic microwave background (CMB) in their canonical version. For instance, the non-canonical version of the chaotic inflationary potential, $V(\phi) \sim \lambda\phi^4$, is found to agree with observations for values of $\lambda$ as large as unity ! The exponential potential can also provide a reasonable fit to CMB observations. A central result of this paper is that {\em steep potentials} (such as $V \propto \phi^{-n}$) usually associated with dark energy, can drive inflation in the non-canonical setting. Interestingly, non-canonical scalars violate the consistency relation $r = -8n_T$, which emerges as a {\em smoking gun} test for this class of models.
}
\keywords{Inflation}
\begin{document}
\maketitle

\section{Introduction}
\label{sec:intro}

Since its inception over three decades ago \cite{models} the inflationary paradigm has moved
to the center stage of modern cosmology.
The amelioration of the horizon and flatness problems --
a generic feature of the inflationary mechanism -- was spectacularly confirmed
by cosmic microwave background (CMB) experiments of the 1990's.
The other important prediction of inflation -- that of a near scale invariant
perturbation spectral index \cite{pert} has also found convincing support in recent
CMB measurements \cite{wmap}.
Yet other important consequences of the inflationary scenario, such as the presence of
a relic gravity wave background and possible departures from non-Gaussianity, will
be tested by current and future probes of the CMB including the PLANCK surveyor, QUaD, BICEP2,
SPIDER, CMBpol, etc.

Despite its remarkable successes the inflationary scenario has, on occasion,
been described as a paradigm in search of a model \cite{turner}.
While models of inflation abound in the literature
\cite{linde_book,ll00,Riotto-2002,Martin-2004,Bassett-2005,Kinney-2009,Sriram-2009,Baumann-2009}, the simplest inflationary
potential described by the quartic self-interaction $\lambda\phi^4$ runs into trouble
with the CMB. %both because of a large gravity wave contribution
The problems with this potential are two fold: (i) its prediction of tensor fluctuations
(gravity waves) are too large and appear to conflict with current bounds on the
tensor-to-scalar ratio. (ii) The value of the dimensionless constant, $\lambda$,
inferred from CMB observations is anomalously small, $\lambda \sim 10^{-13}$,
much smaller for instance than the coupling constant of the Higgs boson
\cite{pdb} $(\lambda \sim 0.1)$.

Perhaps the simplest generalization of the inflationary scenario involves extending
the inflationary Lagrangian to accommodate non-canonical kinetic terms.
It is well known that the equations of motion remain second order, which is an
attractive feature of this class of models.
Furthermore, as we demonstrate in this paper, the slow-roll conditions become
easier to satisfy with the result that the tensor-to-scalar ratio drops considerably
(relative to the canonical case).
These properties of non-canonical scalars have wide reaching ramifications:

\begin{itemize}

\item
Chaotic inflation with the $\lambda\phi^4$ potential
satisfies current CMB constraints, and large %and more physically appealing
values of $\lambda \sim O(1)$ can be accommodated by the data making this
model physically appealing.
(By comparison an astonishingly small value
 $\lambda \sim 10^{-13}$ is demanded of canonical inflation by observations.)

\item
The exponential potential, which is in tension with observations in the canonical
case, also comes into favor since its tensor-to-scalar ratio can lie within
the observationally acceptable range.
As a result the potential $V = V_0[\cosh{(\lambda\phi}) - 1]$ which allows an
exponential-type potential to oscillate
becomes an interesting inflationary contender.

\item
Steep potentials such as $V \propto \phi^{-n}$, which are commonly associated with
dark energy in the canonical case, can source inflation for non-canonical fields.

\item
As pointed out in \cite{Garriga-1999}, inflation sourced by non-canonical scalars violates the
consistency relation $r = -8n_T$, which emerges as a {\em smoking gun}
for this class of models.

\end{itemize}

One might add that the purpose of the present paper is not to add yet another
theoretical construct to the already burgeoning inflationary-model inventory.
Instead, building on earlier work \cite{Picon-1999,Garriga-1999},
 we present a new dynamical framework flexible
enough to accommodate within its fold different classes of inflationary models
including large field models such as chaotic inflation, as well as small field models.
Indeed the formulae presented in \S \ref{sec:cosmology}  are sufficiently general to allow
the reader to translate a canonical inflationary model into its non-canonical
counterpart, obtaining in the process important observational quantities including
power spectra, the spectral indices $n_s$, $n_T$, the tensor-to-scalar ratio,
$r$, etc.

This paper is organized as follows.
The field equations are set up in section \ref{sec:cosmology} which also contains a discussion
of the slow roll parameters and the Hamilton-Jacobi equation for non-canonical scalars.
Inflationary models and CMB constraints on model parameters
 are the focus of section \ref{sec:inflation}.
Non-canonical scalar fields have difficulty in oscillating
which could make
reheating problematic in this scenario. This important drawback is both noticed and corrected
in section \ref{sec:motion}. Our main conclusions are drawn in section \ref{sec:conclusions}.

\section{Cosmological dynamics}
\label{sec:cosmology}

\subsection{Field equations}
\label{sec:1}
%Our model of inflation will be based on the non-canonical scalar field
%Lagrangian
Consider a scalar field which couples minimally to gravity and for which the action has
 the following general form
\beq
S[\phi]=\int\!\d^{4}x\, \sqrt{-g}\; {\cal L}(X,\phi),\label{eqn: action}
\eeq
where
 the Lagrangian density ${\cal L}(\phi , X)$
can be an arbitrary function of the field $\phi$ and the
kinetic term
\beq
X=\frac{1}{2}\pa_{\mu}\phi\; \pa^{\mu}\phi.\label{eqn: X-phi}
\eeq
Numerous functional forms of ${\cal L}(\phi , X)$ have been considered in the literature, see for instance Refs.\cite{Picon-1999,Garriga-1999,Fairbairn-2002,Steer-2004,Campo-2009,sanil-2008,
Mukhanov-2006,Panotopoulos-2007,Chimento-2004, udmde, kdq, Picon-2003, haloes-k}.
Varying the action (\ref{eqn: action}) with respect to $\phi$ leads to the
equation of motion
\ber
\frac{\pa {\cal L}}{\pa \phi} - \l(\frac{1}{\sqrt{-g}}\r)\pa_{\mu}\l(\sqrt{-g}\frac{\pa {\cal L}}{\pa \l(\pa_{\mu}\phi\r)}\r) = 0.\label{eqn: EOM1}
\eer
In a spatially flat Friedmann-Robertson-Walker (FRW) universe
\beq
\d s^2 = \d t^2-a^{2}(t)\; \l[\d x^2 + \d y^2 + \d z^2\r],
\label{eqn: FRW}
\eeq
the field $\phi$ is a function only of time \emph{i.e.}, $\phi = \phi(t)$,
hence the equation of motion (\ref{eqn: EOM1}) reduces to
\beq
\l[\l(\f{\pa {\cal L}}{\pa X}\r)\,
+\, \l(2\, X\r)\, \l(\f{\pa^{2}{\cal L}}{\pa X^{2}}\r)\r]\,
{\ddot \phi}\;+\; \l[\l(3\, H\r)\, \l(\f{\pa {\cal L}}{\pa X}\r)\,
+\, {\dot \phi}\, \l(\f{\pa^{2}{\cal L}}{\pa X\, \pa\phi}\r)\r]\,
{\dot \phi}\;
-\; \l(\f{\pa {\cal L}}{\pa \phi}\r)=0,\label{eqn: EOM-FRW}
\eeq

The energy-momentum tensor
associated with the scalar field is
\beq
T^{\mu\nu}
=\l(\f{\pa{\cal L}}{\pa X}\r)\, \l(\pa^{\mu}\phi\; \pa^{\nu}\phi\r)
- g^{\mu\nu}\, {\cal L}~.\label{eqn: SET}
\eeq
In a spatially flat FRW universe
\beq
T^{\mu}_{\;\:\;\nu} = \mathrm{diag}\l(\rho_{_{\phi}}, -p_{_{\phi}}, - p_{_{\phi}}, - p_{_{\phi}}\r),
\eeq
where the energy density, $\rho_{_{\phi}}$, and pressure, $p_{_{\phi}}$, are given by
\ber
\rho_{_{\phi}} &=& \l(\f{\pa {\cal L}}{\pa X}\r)\, (2\, X)- {\cal L}\label{eqn: rho-phi},\\
p_{_{\phi}} &=& {\cal L}\label{eqn: p-phi},
\eer
and  $X = ({\dot \phi}^{2}/2)$.
The evolution of the scale factor $a(t)$ is governed by the Friedmann equations:
\ber
\l(\frac{\dot{a}}{a}\r)^{2} &=& \l(\frac{8 \pi G}{3}\r)\rho_{_{\phi}},\label{eqn: Friedmann eqn1}\\
\frac{\ddot{a}}{a} &=& -\l(\frac{4 \pi G}{3}\r)\l(\rho_{_{\phi}} + 3\,p_{_{\phi}}\r).\label{eqn: Friedmann eqn2}
\eer
where $\rho_{_{\phi}}$ satisfies the conservation equation
\beq
{\dot \rho_{_{\phi}}} = -3\, H \l(\rho_{_{\phi}} + p_{_{\phi}}\r), ~~ H \equiv \frac{\dot a}{a}~.
\label{eqn: conservation eqn}
\eeq
Note that the  equation of motion for $\phi$  in (\ref{eqn: EOM-FRW})
also follows from the conservation equation (\ref{eqn: conservation eqn}).

Our non-canonical scalar field model has the Lagrangian density \cite{sanil-2008,Mukhanov-2006}
\beq
{\cal L}(X,\phi) = X\l(\frac{X}{M^{4}}\r)^{\alpha-1} -\; V(\phi),
\label{eqn: Lagrangian}
\eeq
where $M$ has dimensions
 of mass while $\alpha$ is dimensionless. When $\alpha = 1$
the Lagrangian (\ref{eqn: Lagrangian}) reduces to the usual canonical scalar field Lagrangian
${\cal L}(X,\phi) = X -\; V(\phi)$.
 Throughout this paper
 we shall assume $\mpl = 1/\sqrt{8\pi G}$ and work with natural units, \emph{viz}.\
$c = \hbar \equiv 1$.

The energy density and pressure are obtained by substituting (\ref{eqn: Lagrangian}) into
(\ref{eqn: rho-phi}) and (\ref{eqn: p-phi}), we find
\ber
\rho_{_{\phi}} &=& \l(2\alpha-1\r)X\l(\frac{X}{M^{4}}\r)^{\alpha-1} +\;  V(\phi),\nonumber\\
p_{_{\phi}} &=& X\l(\frac{X}{M^{4}}\r)^{\alpha-1} -\; V(\phi), ~~
X \equiv \frac{1}{2} {\dot \phi}^{2}~,
\label{eqn: p-model}
\eer
which reduces to the canonical form
$\rho_{_{\phi}} = X + V$, ~$p_{_{\phi}} = X - V$ when $\alpha = 1$.
Consequently
the two Friedmann equations (\ref{eqn: Friedmann eqn1}) and (\ref{eqn: Friedmann eqn2}) become
\cite{Mukhanov-2006,Panotopoulos-2007}
\ber
H^{2} &=& \frac{8 \pi G}{3}\l[\l(2\alpha-1\r)X\l(\frac{X}{M^{4}}\r)^{\alpha-1} +\;
V(\phi)\r]~,\label{eqn: FR-eqn1 model}\\
\frac{\ddot{a}}{a} &=& -\frac{8 \pi G}{3}\l[\l(\alpha + 1\r)X\l(\frac{X}{M^{4}}\r)^{\alpha-1} -\;  V(\phi)\r]~,
\label{eqn: FR-eqn2 model}
\eer
and the following scalar field
 equation of motion follows from eq.(\ref{eqn: EOM-FRW})
\beq
{\ddot \phi}+ \f{3\, H{\dot \phi}}{2\alpha -1} + \l(\f{V'(\phi)}{\alpha(2\alpha -1)}\r)\l(\f{2\,M^{4}}{{\dot \phi}^{2}}\r)^{\alpha - 1} =\; 0,
\label{eqn: EOM-model}
\eeq
%where $V'(\phi) = \frac{\d V(\phi)}{\d\,\phi}$.
%Equation (\ref{eqn: EOM-model})
which reduces to
${\ddot \phi}+ 3\, H {\dot \phi} + V'(\phi) = 0$ when $\alpha =1$.

\subsection{Slow roll parameters}
\label{sec:slow roll}

The two slow roll parameters are defined according to convention as
\beq
\varepsilon \equiv -\frac{\dot{H}}{\,H^{2}}~, ~~
\delta \equiv \varepsilon -  \frac{\dot{\varepsilon}}{2\,H\,\varepsilon}~.
\label{eqn: slow roll}
\eeq
Since
$\ddot{a}/{a} = 1 - \varepsilon$, it follows that the universe accelerates
(inflates) when $\varepsilon < 1$. % and inflation ends when $\varepsilon \simeq 1$.
 Substituting for $\rho_{_{\phi}}$ and $p_{_{\phi}}$ from (\ref{eqn: p-model}) into
\beq
\dot{H} = -4\pi G \l(\rho_{_{\phi}} + p_{_{\phi}}\r)
= - 8\pi G\alpha X\left (\frac{X}{M^4}\right )^{\alpha-1}~,
\eeq
it is easy to show that the FRW equation (\ref{eqn: FR-eqn1 model}) reduces to
\beq
H^{2}\l[1 - \l(\frac{2\alpha -1}{3\alpha}\r)\varepsilon\r] = \frac{8 \pi G}{3} V(\phi)~.
\label{eqn: H-epsilon}
\eeq
As concerns the second slow roll parameter, one finds
\beq
\delta = -\alpha\l(\frac{\ddot{\phi}}{H\dot{\phi}}\r) ~,\label{eqn: delta-phidd}
\eeq
and substituting (\ref{eqn: delta-phidd}) into (\ref{eqn: EOM-model}) we obtain
\beq
3\,H\,{\dot \phi}\l[1 - \l(\frac{2\alpha -1}{3\alpha}\r)\delta\r] =
- \f{V'}{\alpha}\l(\f{2\,M^{4}}{{\dot \phi}^{2}}\r)^{\alpha - 1}~.
\label{eqn: phidot-delta}
\eeq
It is easy to show that the slow roll conditions $\varepsilon \ll 1$, $|\delta| \ll 1$, imply
the following relations between the slow roll parameters and the inflaton potential:
\beq
\varepsilon \simeq \varepsilon_{_V} = \l[\frac{1}{\alpha}
\l(\frac{3\,M^{4}}{V}\r)^{\alpha -1}
\l(\frac{\mpl\,V'}{\sqrt{2}\;V}\r)^{2\alpha}\r]^{\frac{1}{2\alpha - 1}},
\label{eqn: potential SR1}
\eeq
\beq
\delta \simeq \l(\frac{\alpha \varepsilon}{2\alpha-1}\r)\l(2\Gamma - 1\r)~,
\label{eqn: potential SR2}
\eeq
where the parameter
\beq
\Gamma = \frac{V(\phi)V''(\phi)}{V'(\phi)^{2}}
\eeq
plays a key role in inflationary and quintessence model building.
For the canonical scalar field with $\alpha = 1$ equations
(\ref{eqn: potential SR1}), (\ref{eqn: potential SR2}) converge to
the standard `canonical' expressions
\ber
\varepsilon_{_V}^{(c)} &=& \frac{\mpl^{2}}{2}\l(\frac{V'}{V}\r)^{2}~,\label{eq:standard_slowroll}\\
\delta^{(c)} &=& %\varepsilon \l(2\Gamma - 1\r)
\mpl^{2}\l(\frac{V''}{V} - \frac{1}{2}\l(\frac{V'}{V}\r)^2\r)~.
\label{eq:standard_slowroll1}
\eer
Note that the slow roll assumption leads to
\beq
{\dot \phi} = -\theta\l\{\l(\frac{\mpl}{\alpha\,\sqrt{3}}\r)\l(\frac{\theta\,V'(\phi)}{\sqrt{V}}\r)
\l(2\,M^{4}\r)^{\alpha-1}\r\}^{\frac{1}{2\alpha-1}},\label{eqn: phidot}
\eeq
where
$\theta = +1$ when $V'(\phi) > 0$;
$\theta = -1 $ when $V'(\phi) < 0.$
This equation shall prove useful when we derive an expression for the number of
inflationary e-folds in the next section.

Equation~(\ref{eqn: potential SR1}) can be written in a more suggestive manner as
\beq
\varepsilon_{_V} = \l(\frac{1}{\alpha}\r)^{\frac{1}{2\alpha - 1}}
\l(\frac{3\,M^{4}}{V}\r)^\frac{\alpha -1}{2\alpha - 1}
\left [\varepsilon_{_V}^{(c)}\right ]^\frac{\alpha}{2\alpha-1}%{\bigg\vert_{\alpha=1}}
\label{eq:epsilon}
\eeq
where $\varepsilon_{_V}^{(c)}$ corresponds to the canonical value of $\varepsilon_{_V}$
in (\ref{eq:standard_slowroll}).
Since $\alpha > 1$ it follows that the first term in the right hand side of (\ref{eq:epsilon}) is smaller
than unity.
%since $2\alpha/(2\alpha -1) > 1$, it follows that for values of the parameter
We therefore find that the slow roll parameter in non-canonical models can become
smaller than its canonical counterpart, \ie ~ $\varepsilon_{_V} < \varepsilon_{_V}^{(c)}$,
when $3M^4 \ll V$.
We shall return to this issue in section \ref{sec:inflation}, where we show that sub-Planckian values of
$M$ provide better
agreement with observations for a large family of inflationary models.
Another aspect of (\ref{eq:epsilon}) is that inflation can be sourced by steep potentials
in non-canonical models, which is discussed next.

\subsection{Inflation with steep potentials}

Equation (\ref{eq:epsilon}) with $\alpha > 1$ implies the inequality
\beq
\varepsilon_{_V} < \l(\frac{3\,M^{4}}{V}\r)^\frac{\alpha -1}{2\alpha - 1}
\big [\varepsilon_{_V}^{(c)}\big ]^\frac{\alpha}{2\alpha-1}%{\bigg\vert_{\alpha=1}}
\label{eq:epsilon1}
\eeq
which allows inflation to be sourced by steep potentials when $V \gg M^{4}$.
Note that the possibility of sourcing inflation using steep potentials has earlier been
discussed in the braneworld context in \cite{maartens}.

Indeed, equation (\ref{eq:epsilon1}) bears a close similarity to the relationship between slow
roll parameters in an RSII braneworld cosmology. The latter is described by the equations
\cite{rs,brane,maartens}
\beq
H^2 = \frac{1}{3 \mpl^2}\rho \l(1 + \frac{\rho}{2\lambda_b}\r),
\label{eq:frw2}
\eeq
where $\rho = \frac{1}{2}{\dot\phi}^2 + V(\phi)$ and $\lambda_b$ is the three dimensional brane tension.
As noted by a number of authors \cite{maartens,copeland,sss02}
the motion of a canonical scalar field propagating on the brane
\begin{equation}
{\ddot \phi} + 3H {\dot \phi} + V'(\phi) = 0~,
\label{eq:kg}
\end{equation}
is heavily damped due to the increased value of the term $3H{\dot\phi}$ when
$\rho/\lambda_b \gg 1$ in (\ref{eq:frw2}).
This causes the field to roll slower than it would in an FRW cosmology which is reflected in
%the following relationship between the value of the slow roll parameter on the brane
the value of the slow roll parameter on the brane
%$\varepsilon_b$ and its canonical value $\varepsilon_{_V}^{(c)}$ \cite{sss02}
\beq
\varepsilon_b \simeq \left (\frac{4\lambda_b}{V}\right )\varepsilon_{_V}^{(c)}~, ~~ V/\lambda_b \gg 1
\label{eq:slow-rollbrane}
\eeq
where $\varepsilon_{_V}^{(c)}$ is the canonical value in an FRW
universe, namely (\ref{eq:standard_slowroll1}).
%Clearly $\varepsilon_b \ll \varepsilon_{_V}^{(c)}$ when $V \gg \lambda_b$.
Comparing (\ref{eq:epsilon1}) \& (\ref{eq:slow-rollbrane}) we find that the parameter $M$
in non-canonical models plays a role similar to that of the brane tension $\lambda_b$
in braneworld cosmology.
In the braneworld case $\varepsilon_b \ll \varepsilon_{_V}^{(c)}$ when $V \gg \lambda_b$
in (\ref{eq:slow-rollbrane}),
while for non-canonical scalars
$\varepsilon_{_V} \ll \varepsilon_{_V}^{(c)}$ when $V \gg M^{4}$ in (\ref{eq:epsilon}) or (\ref{eq:epsilon1}).

It is well known that inflation can be driven by steep potentials
in braneworld models \cite{maartens,copeland,sss02} and this intriguing possibility can be
realized for non-canonical scalars as well.
We illustrate this for two potentials commonly associated with dark energy
in the canonical case:
(i) the inverse power law potential \cite{ratra88} $V \propto \phi^{-n}$ and (ii) the exponential potential
\cite{zws} $V \propto (\exp{\frac{\mpl}{\phi}} - 1)$.

\begin{enumerate}

\item
%Note in passing that $\varepsilon_{_V}$ evaluated for the inverse power law potential
For the inverse power law potential
\beq
V = \frac{{\cal M}^4}{(\phi/\mpl)^n}
\label{eq:IPL}
\eeq
%gives the following result
one finds
\beq
\varepsilon_{_V} = \left (\frac{\phi}{\mpl}\right )^\frac{n(\alpha-1)-2\alpha}{2\alpha-1}\left [\frac{n^{2\alpha}}{\alpha 2^\alpha}\left (\frac{3M^4}{{\cal M}^4}
\right )^{\alpha - 1}\right ]^\frac{1}{2\alpha-1}
\label{eq:inversePL}
\eeq
so that for $n = 2\alpha/(\alpha - 1)$
%\beq
%n = \frac{2\alpha}{\alpha - 1}
%\eeq
the value of $\varepsilon_{_V}$ {\em does not depend} upon $\phi$ !
In this case
\beq
\varepsilon_{_V} = \left [\frac{n^{2\alpha}}{\alpha 2^\alpha}\left (\frac{3M^4}{{\cal M}^4}
\right )^{\alpha - 1}\right ]^\frac{1}{2\alpha-1}
\eeq
and it is easy to see that $\varepsilon_{_V} < 1$ for ${\cal M} \gg M$ and $\alpha > 1$.
We therefore find that the inverse power law potential, which is associated with
dark energy in its canonical version, can source inflation for non-canonical fields !
From (\ref{eq:inversePL}) we also find that
\beq
\varepsilon_{_V} \ll \left [\frac{n^{2\alpha}}{\alpha 2^\alpha}\left (\frac{3M^4}{{\cal M}^4}
\right )^{\alpha - 1}\right ]^\frac{1}{2\alpha-1}
\eeq
 when $\phi \ll \mpl$ and $n > 2\alpha/(\alpha - 1)$. Therefore we arrive at the following
interesting result: in a non-canonical setting it may be possible for
the inverse power law potential (\ref{eq:IPL}) to source
{\em small field inflation} provided $n \geq 2\alpha/(\alpha - 1)$ and ${\cal M} \gg M$.

\item For the exponential potential
\beq
V(\phi) = V_0\left (e^{\mpl/\phi} - 1\right )
\label{eq:expo}
\eeq
one finds%, when $\phi \ll \mpl$, $V \propto e^{\mpl/\phi}$ and
\beq
\varepsilon_{_V}^{(c)} = \frac{1}{2}\left (\frac{\mpl}{\phi}\right )^4 \gg 1~, ~~ {\rm for} ~~\phi \ll \mpl
\eeq
which rules out canonical inflation for such ultra-steep potentials.

However the non-canonical slow roll parameter (\ref{eq:epsilon1}) acquires the form
\beq
\varepsilon_{_V} < \left [ \frac{3M^4}{V_0}e^{-\mpl/\phi}\right ]^\frac{\alpha - 1}{2\alpha - 1}
\left (\frac{\mpl}{\phi}\right )^\frac{4\alpha}{2\alpha - 1}
\label{eq:expo1}
\eeq
and one sees that $\varepsilon_{_V} \ll 1$ is  easily achievable for $\phi \ll \mpl$
and $V_0 > M^4$ since the exceedingly small value of the exponential term in the RHS of
(\ref{eq:expo1}) easily compensates the large value of the $\mpl/\phi$ term.

\end{enumerate}

We therefore conclude that non-canonical models bring more diversity into inflationary model
building by permitting inflation to be sourced by flat as well as steep potentials.

As in the braneworld case \cite{copeland,sss02,liddle10}, the possibility of sourcing inflation using steep potentials
might allow one to construct models of {\em Quintessential Inflation} \cite{pv99} based on non-canonical
scalars.
We shall return to this possibility in a future work.

We end this section with a cautionary note.
It is formally not very meaningful to claim that a potential is steep without
reference to the kinetic term. For instance, one can
convert a flat potential into a steep one using
a field redefinition, but then one also changes the form of the kinetic term,
as illustrated in the appendix.
 For this reason
 one should exercise some care when comparing
models with different kinetic terms.
In our discussion above, results for the braneworld model were based on
the behavior of a canonical scalar field propagating on the brane, whereas our own
model is based on the non-canonical Lagrangian (\ref{eqn: Lagrangian}).
(The propagation of non-canonical scalar fields on the brane has, to the best of our
knowledge, not yet been studied.)

\subsection{Hamilton-Jacobi equation and the Inflationary Attractor}
\label{sec:hamilton}

The slow roll parameters on their own do not necessarily encapsulate the full dynamical
picture of inflation. The fact that the equations of motion of the inflaton are of second
order makes it possible to change initial conditions (value of ${\dot\phi}$)
so as to arrive at a different set of observational predictions for inflation \cite{ll00}.
Clearly in order for inflation to be a robust theory its predictions should not
be unduly sensitive to initial conditions.
In other words, it would be desirable if the difference between nearby trajectories were
to rapidly decay during inflation. That this is indeed the case was shown in several
early papers which demonstrated the existence of an inflationary attractor solution
\cite{belinsky}.

The presence of the inflationary attractor is easiest to demonstrate using the Hamilton-Jacobi
formalism \cite{sb90}, and we shall adopt this method for our present analysis.
The idea behind the Hamilton-Jacobi formalism is to rewrite the
Friedmann equation (\ref{eqn: FR-eqn1 model}) as
an evolution equation for $H(\phi)$.
For the non-canonical case this is done by noting that
\beq
\frac{\d H}{\d\phi} = \frac{\dot{H}}{\dot{\phi}}
= -4\,\pi G\l(\frac{\rho_{_{\phi}} + p_{_{\phi}}}{\dot{\phi}}\r)~,
\eeq
substituting for $\rho_{_{\phi}}$ and $p_{_{\phi}}$ from
(\ref{eqn: p-model})) and rearranging, gives
\beq
\dot{\phi} = \pm\,\mpl^{2}\l\{\l(\frac{2^{\alpha}\mu^{4(\alpha-1)}}{\alpha}\r)\l(\mp\,H'(\phi)\r)
\r\}^{\frac{1}{2\alpha-1}},\label{eqn: phi dot H}
\eeq
where $$\mu \equiv \frac{M}{\mpl}~.$$
In the above equation overprime denotes derivative with respect to $\phi$.
It is evident from the above equation that the sign of $\dot{\phi}$ and $H'(\phi)$ are opposite to each other.
The above equation implies
\beq
X\l(\frac{X}{M^{4}}\r)^{\alpha-1} = \mpl^{4}\l(\frac{2^{\alpha}\mu^{4(\alpha-1)}}{\alpha^{2\alpha}}\r)^{\frac{1}{2\alpha-1}}
\l[H'(\phi)\r]^{\frac{2\alpha}{2\alpha-1}}
\eeq
on substituting this equation in the  Friedmann equation (\ref{eqn: FR-eqn1 model}), gives
\beq
\l[H'(\phi)\r]^{\frac{2\alpha}{2\alpha-1}}\, -\, \l(\frac{3\,f_{_1}(\alpha)}{2\,\mpl^{2}}\r)H(\phi) = -\l(\frac{\,f_{_1}(\alpha)}{2\,\mpl^{4}}\r)V(\phi),
\label{eqn: H-J eqn}
\eeq
where
\beq
f_{_1}(\alpha)= \l(\frac{1}{2\alpha - 1}\r)\l(\alpha^{2\alpha}\l(\frac{2}{\mu^{4}}\r)^{\alpha-1}\r)^{\frac{1}{2\alpha-1}}.
\label{eqn: f1}
\eeq
Equation (\ref{eqn: H-J eqn}) is Hamilton-Jacobi equation  corresponding to the
non-canonical Lagrangian (\ref{eqn: Lagrangian}). For $\alpha = 1$, $f_{_1}(\alpha) = 1$
and (\ref{eqn: H-J eqn}) reduces to
the standard Hamilton-Jacobi equation for the canonical scalar field, namely
\beq
H'(\phi) - \frac{3}{2\mpl^{2}}H(\phi) = -\frac{1}{2\mpl^{4}}V(\phi)~.
\eeq

The Hamilton-Jacobi equation (\ref{eqn: H-J eqn}) allows us to determine $H(\phi)$ for a given
$V(\phi)$. Conversely, one can also reconstruct the potential $V(\phi)$ if have prior
knowledge of $H(\phi)$.
It is important to note that, for any given $V(\phi)$, each phase-space
 trajectory $\dot{\phi}(\phi)$ can be mapped to a corresponding $H(\phi)$.
This follows from the Friedmann equation (\ref{eqn: FR-eqn1 model}).
Therefore,
for a given potential $V(\phi)$, let $\dot{\phi}(\phi)$ and
$\dot{\phi}(\phi) + \delta\dot{\phi}(\phi)$ be two nearby phase-space trajectories
 corresponding to the homogeneous solutions $H(\phi)$ and $H(\phi) + \delta H(\phi)$, respectively.
Substituting these into the Hamilton-Jacobi equation (\ref{eqn: H-J eqn}) and
linearizing gives
\beq
\frac{\delta H'(\phi)}{\delta H(\phi)} = \l(\frac{3f_{_1}(\alpha)}{\mpl^{2}}\r)\l(\frac{2\alpha - 1}{2\alpha}\r)\l(\frac{H(\phi)}{\l[H'(\phi)\r]^{\frac{1}{2\alpha-1}}}\r)
\label{eqn: linear HJ eqn}
\eeq
which is easily solved to give
\beq
\delta H(\phi) = \delta H(\phi_{_i})\,\exp\l[\l(\frac{3f_{_1}(\alpha)}{\mpl^{2}}\r)\l(\frac{2\alpha - 1}{2\alpha}\r)
\int^{\phi}_{\phi_{_i}}\l(\frac{H(\phi)}{\l[H'(\phi)\r]^{\frac{1}{2\alpha-1}}}\r)\d\phi\r]
\label{eqn: s1}
\eeq
here $\delta H(\phi_{_i})$ is the value of the perturbed Hubble parameter corresponding to some
initial $\phi_{_i}$.
Since, $N$ is number of efolds counted from the end of inflation
\beq
N - N_{_i}= -\int^{\phi}_{\phi_{_{i}}}\l(\frac{H}{\dot{\phi}}\r)\d\,\phi ~,
\eeq
substituting for $\dot{\phi}$ from (\ref{eqn: phi dot H}) into the above equation gives
\beq
\int^{\phi}_{\phi_{_i}}\l(\frac{H(\phi)}{\l[H'(\phi)\r]^{\frac{1}{2\alpha-1}}}\r)\d\phi\,
=\, \mpl^{2}\l(\frac{2^{\alpha}\mu^{4(\alpha-1)}}{\alpha}\r)^{\frac{1}{2\alpha-1}}\l(N - N_{_i}\r)
\label{eqn: s2}
\eeq
Finally, using (\ref{eqn: f1}), (\ref{eqn: s1}) \&  (\ref{eqn: s2}), we find the following
simple expression describing the decay of perturbations
\beq
\delta H(\phi) = \delta H(\phi_{_i})\exp\l[\,-\,3\,(N_{_i}\,-\,N)\r]~,
\label{eqn: HJ soln}
\eeq
and signifying an exponentially rapid approach to the inflationary attractor solution.
Remarkably our final expression (\ref{eqn: HJ soln})
does not depend on either $\alpha$ or $M$, the two
free parameters which characterize our model (\ref{eqn: Lagrangian}). Since
the rate at which a nearby trajectory $H(\phi) + \delta H(\phi)$ converges to $H(\phi)$ is
independent of $\alpha$ and $M$ we conclude
that homogeneous perturbations in non-canonical inflationary models decay in
{\em precisely the same manner} as they do for canonical scalars \cite{sb90}.\\
%the approach to a
%late time inflationary attractor solution can be deemed to be robust.

\leftline{\bf Evolution of slow roll parameter}
%\label{sec:SR-evolution}

The above argument shows that two nearby phase-space trajectories,
whether inflationary or not, converge, thereby ascertaining the dynamical stability of the system.
However, as far as inflation is concerned, it is important to examine whether the inflationary trajectory is an attractor.
To address this issue, we investigate how the slow roll parameter $\varepsilon$ defined in Eq.(\ref{eqn: slow roll}) evolves as the scalar field rolls down the potential.

From the definition of the slow roll parameter (\ref{eqn: slow roll}) and using (\ref{eqn: FR-eqn1 model}) and (\ref{eqn: FR-eqn2 model}), we have
\beq
\varepsilon = \l(\frac{3}{2}\r)\l(\frac{2\alpha\,X\l(X/M^{4}\r)^{\alpha-1}}{\l(2\alpha-1\r)X\l(X/M^{4}\r)^{\alpha-1} +\;  V(\phi)}\r)\label{eqn: ep1}.
\eeq
Therefore, for  $V(\phi) > 0$, the allowed range of $\varepsilon$ is
\beq
0\, \leq\, \varepsilon\, \leq\, \l(\frac{3\,\alpha}{2\alpha-1}\r),
\eeq
where the lower limit corresponds to the equation of state parameter $w = -1$, whereas the upper bound corresponds to
$w = (2\alpha-1)^{-1}$.

Using Eqs.(\ref{eqn: p-model}) and (\ref{eqn: ep1}), the potential $V(\phi)$ can be expressed as
\beq
V(\phi) = \l(\frac{3 \alpha - \l(2\alpha-1\r)\varepsilon}{3 \alpha}\r)\rho_{_{\phi}}.\nn
\eeq
Therefore,
\beq
\frac{V'(\phi)}{V(\phi)} = - \l(\frac{1}{\dot{\phi}}\r)\l(\frac{\l(2\alpha-1\r)\dot{\varepsilon} + \l[3 \alpha - \l(2\alpha-1\r)\varepsilon\r]2\,H\varepsilon}{3 \alpha - \l(2\alpha-1\r)\varepsilon}\r)\label{eqn: vpbyv}
\eeq
Since $\varepsilon \equiv -\dot{H}/H^{2}$, it follows that $H'(\phi) = -\varepsilon H^{2}/\dot{\phi}$ and therefore,
Eq.(\ref{eqn: phi dot H}) can be re-expressed as
\beq
\dot{\phi} = -\,\theta\l(\frac{\varepsilon\,2^{\alpha}\,\mpl^{2}\,H^{2}\,M^{4(\alpha -1)}}{\alpha}\r)^{\frac{1}{2\alpha}}
\label{eqn: phidotH}
\eeq
where
$\theta = +1$ when $V'(\phi) > 0$ and
$\theta = -1 $ when $V'(\phi) < 0.$
The sign of $\theta$ ensures that scalar field rolls down the potential.
Substituting Eq.(\ref{eqn: phidotH}) in Eq.(\ref{eqn: vpbyv}) and on rearranging, we arrive at the following equation of motion for $\varepsilon$:
\beq
\dot{\varepsilon}\, =\, -\,2\,H\,\varepsilon\,\l(\frac{3\alpha}{2\alpha -1}-\varepsilon\r)
\l[1\, - \, \l(\frac{V(\phi)}{3H^{2}\mpl^{2}}\r)^{\frac{\alpha -1}{2\alpha}}
\l(\frac{\varepsilon_{_V}}{\varepsilon}\r)^{\frac{2\alpha -1}{2\alpha}}\r],
\label{eqn: EOM-epsilon}
\eeq
where $\varepsilon_{_V}$ is the slow roll parameter defined in terms of the potential in
(\ref{eqn: potential SR1}).

Note that equation (\ref{eqn: EOM-epsilon}) is exact and the slow roll approximation has not been
 used in its derivation.
Using (\ref{eqn: EOM-epsilon}), one can investigate how the slow roll parameter $\varepsilon$ evolves as the
scalar field rolls down its potential and thereby
 ascertain whether the slow roll inflationary trajectory ($\varepsilon \simeq \varepsilon_{_V}$)
is an attractor.
Let us first consider the canonical scalar field case which corresponds to setting $\alpha = 1$ in
(\ref{eqn: EOM-epsilon}) so that this equation
reduces to
\beq
\dot{\varepsilon}\, =\, -\,2\,H\,\varepsilon\,\l(3\,-\,\varepsilon\r)
\l(1\, - \, \sqrt{\frac{\varepsilon_{_V}}{\varepsilon}}\;\r),
\label{eqn: EOM-epsilon-C}
\eeq
where $\varepsilon_{_V}$ is given by (\ref{eq:standard_slowroll}) for canonical inflation.
Since $\varepsilon$ always lies between $0$ and $3$ and $H > 0$ in an expanding universe,
the sign of $\dot{\varepsilon}$ (at a given value of $\phi$) is determined by the
last term in the left hand side of (\ref{eqn: EOM-epsilon-C}).
In other words the sign of $\dot{\varepsilon}$ depends upon the value of ${\varepsilon_{_V}}/{\varepsilon}$,
 which quantifies the departure  of $\varepsilon$ from its slow roll value $\varepsilon_{_V}$.
\begin{itemize}
\item[$\ast$] Eqn (\ref{eqn: EOM-epsilon-C}) suggests that if, for a given phase-space value
$\lbrace \phi, \dot{\phi}\rbrace$ one finds
$\varepsilon\, >\, \varepsilon_{_V}$, then
$\dot{\varepsilon}\, <\, 0$ so that $\varepsilon$ is driven towards $\varepsilon_{_V}$.
  \item[$\ast$] On the other hand if $\varepsilon\, <\, \varepsilon_{_V}$, then $\dot{\varepsilon}\, >\, 0$,
so in this case also $\varepsilon$ is driven towards $\varepsilon_{_V}$.
%\item[$\ast$] Therefore, \textbf{in the regime where $0\, < \varepsilon_{_V}\, < 1$, different phase trajectories evolves such that $\varepsilon$ is driven towards $\varepsilon_{_V}$.
%Hence slow roll inflationary trajectory where $\varepsilon_{_V}\, \ll 1$ is actually an attractor.}
\end{itemize}
We therefore find that $\varepsilon$ is always driven towards $\varepsilon_{_V}$
so that the slow roll trajectory ($\varepsilon_{_V}\, \ll 1$) is an attractor.

It is easy to see that similar results follow from (\ref{eqn: EOM-epsilon}) for $\alpha \neq 1$.
The slow roll inflationary trajectory corresponds to $3\mpl^{2}H^{2} \simeq V(\phi)$ and
$\varepsilon \simeq \varepsilon_{_V}$.
If, for
a given value of $\phi$, $\varepsilon_{_V}\, \ll 1$, and the value of $\dot{\phi}$ is such that:
\begin{itemize}
\item[$\ast$] $\varepsilon\, >\, \varepsilon_{_V}$, then it follows from
(\ref{eqn: EOM-epsilon}) that
$\dot{\varepsilon}\, <\, 0$ and $\varepsilon$ is driven towards $\varepsilon_{_V}$.
\item[$\ast$] $\varepsilon\, <\, \varepsilon_{_V}$, then $\dot{\varepsilon}\, >\, 0$
is implied by (\ref{eqn: EOM-epsilon}),  so that once more
 $\varepsilon$ is driven towards $\varepsilon_{_V}$.
\end{itemize}
We therefore conclude  that for any potential possessing
 a regime satisfying $\varepsilon_{_V}\, \ll 1$,
the slow roll inflationary trajectory is indeed an attractor.

Consider next an inflationary potential for which the slow roll parameter in
(\ref{eqn: EOM-epsilon}) becomes
very small, \ie~ $\varepsilon_{_V}\, \ll 1$.
The worst case scenario for inflation is clearly when the kinetic term is very large
$({\dot\phi}^2 \gg V)$ so that $\varepsilon \sim 1$
and $\varepsilon \gg \varepsilon_{_V}$.
%Next note that in the regime where $\varepsilon_{_V}\, \ll 1$, if the value of $\dot{\phi}$ is such that  the slow roll parameter $\varepsilon$ is close to unity, Eq.(\ref{eqn: EOM-epsilon}) can be approximated as:
It is easy to see that in this case (\ref{eqn: EOM-epsilon}) can be approximated as
\beq
\dot{\varepsilon}\, \simeq\, -2\,H\,\varepsilon\l(\frac{3\alpha}{2\alpha -1}-\varepsilon\r),
\eeq
which has the solution
\beq
\varepsilon(a) \simeq\, \l(\frac{3\alpha}{2\alpha -1}\r)
\l[1 + \l(\frac{a\,}{a_{\star}}\r)^{\frac{6\alpha}{2\alpha -1}}\r]^{-1}~,
\label{eqn: soln-epsilon}
\eeq
where $a_{\star}$ is a constant of integration.
From (\ref{eqn: soln-epsilon}) it follows
  that as the universe expands ($a > a_{\star}$) the slow roll parameter
$\varepsilon$ decays as $\varepsilon \propto  a^{-6\alpha/(2\alpha -1)}$ and
soon approaches $\varepsilon_{_V}$, signalling the advent of slow-roll and $w \simeq -1$.

\subsection{Scalar and tensor power spectra}
\label{sec:scalar}

Linearized scalar and tensor perturbations within the spatially flat FRW context are described
by the line element \cite{Bardeen-1980,Kodama-1984,Mukhanov-1992}
\beq
\d s^2
= (1+2\, A)\,\d t ^2 - 2\, a(t)\, (\pa_{i} B )\; \d t\; \d x^i\,
-a^{2}(t)\; \l[(1-2\, \psi)\; \delta _{ij}+ 2\, \l(\pa_{i}\, \pa_{j}E \r) + h_{ij}\r]\,
\d x^i\, \d x^j\nn
\eeq
where $A$, $B$, $\psi$ and $E$ describe
the scalar degree of metric perturbations while $h_{ij}$ are tensor perturbations.
We consider only scalar and tensor perturbations
since it is well known that scalar fields do not lead to vector perturbations.
The curvature perturbation $\mathcal{R}$ on the uniform field slicing
is defined as a gauge invariant combination of the metric perturbation $\psi$
and scalar field perturbation $\delta \phi$, namely
%The expression for the curvature perturbation $\mathcal{R}$ is given by
\beq
\mathcal{R} \equiv \psi + \l(\frac{H}{\dot{\phi}}\r)\delta \phi~.
\eeq
From the Linearized Einstein's equation $\delta G^{\mu}_{\;\nu} = \kappa\, \delta T^{\mu}_{\;\nu}$ and from the equation governing the evolution of perturbations in the scalar field, it turns out that
\beq
\mathcal{R}_{_k}'' + 2\l(\frac{z'}{z}\r)\mathcal{R}_{_k}'  + c_{_s}^{2}k^{2}\mathcal{R}_{_k} = 0~,
\label{eqn: curvature pert eqn}
\eeq
where overprime denotes derivative with respect to conformal time, $\eta = \int dt/a(t)$;
$c_{_s}^{2}$ is the square of the effective speed of sound of the scalar field perturbation~\cite{Garriga-1999}
\beq
c_{_s}^{2} \equiv \l[\f{\l({\pa {\cal L}}/{\pa X}\r)}{\l({\pa {\cal L}}/{\pa X}\r)
+ \l(2\, X\r)\, \l({\pa^{2} {\cal L}}/{\pa X^{2}}\r)}\r] ~,
\label{eq:sound_speed_def}
\eeq
and $z$ is given by
\beq
z \equiv \frac{a\,\l(\rho_{_{\phi}}+ p_{_{\phi}}\r)^{1/2}}{c_{_s}H}.
\eeq

%The Mukhanov-Sasaki variable $u_{_{k}}$ is defined as $u_{_{k}} \equiv z\,\mathcal{R}_{_k}$.
Rewriting (\ref{eqn: curvature pert eqn})
in terms of the Mukhanov-Sasaki variable $u_{_{k}} \equiv z\,\mathcal{R}_{_k}$,
one gets
\beq
u_{_k}'' +  \l(c_{_s}^{2}k^{2} -   \frac{z''}{z}\r)u_{_k} = 0 ~.
\eeq
The corresponding equation governing the tensor perturbations is
\beq
v_{_k}'' +  \l(k^{2} -   \frac{a''}{a}\r)v_{_k} = 0 ~,
\eeq
where $v_{_k} \equiv (h/a)$ and  $h$ is the amplitude of the tensor perturbation.

The power spectrum of scalar curvature perturbations is defined as
\beq
\mathcal{P}_{_{S}}(k) \equiv \l(\frac{k^{3}}{2\pi^{2}}\r)|\mathcal{R}_{_k}|^{2} = \l(\frac{k^{3}}{2\pi^{2}}\r)\l(\frac{|u_{_k}|}{z}\r)^{2} ~,
\eeq
while the tensor power spectrum is
\beq
\mathcal{P}_{_{T}}(k) \equiv 2\l(\frac{k^{3}}{2\pi^{2}}\r)|h_{_k}|^{2} = 2\l(\frac{k^{3}}{2\pi^{2}}\r)\l(\frac{|v_{_k}|}{a}\r)^{2}.
\label{eq:tensor}
\eeq

Following \cite{Garriga-1999}, the expression for scalar and tensor power spectrum in the
slow roll limit turns out to be
\beq
\mathcal{P}_{_{S}}(k) = \l(\frac{H^{2}}{2\pi\l(c_{_s}\l(\rho_{_{\phi}}+ p_{_{\phi}}\r)\r)^{1/2}}\r)^{2}_{aH\, =\, c_{_s}k}
\label{eqn: scalar PS}
\eeq
and
\beq
\mathcal{P}_{_{T}}(k) = \l(\frac{8}{\mpl^{2}}\r)\l(\frac{H}{2\pi}\r)^{2}_{aH\, =\, k}
\simeq \l(\frac{2\,V(\phi)}{3\,\pi^{2}\mpl^{4}}\r)_{aH\, =\, k}~.
\label{eqn: tensor PS}
\eeq
For the Lagrangian density (\ref{eqn: Lagrangian}), the scalar power spectrum
 in the slow roll regime is determined from (\ref{eqn: scalar PS})
to be
\beq
\mathcal{P}_{_{S}}(k) = \l(\frac{1}{72\pi^{2}c_{_s}}\r)\l\{\l(\frac{\alpha\, 6^{\alpha}}{\mu^{4(\alpha-1)}}\r)
\l(\frac{1}{\mpl^{14\alpha -8}}\r)\l(\frac{V(\phi)^{5\alpha - 2}}{V'(\phi)^{2\alpha}}\r)
\r\}^{\frac{1}{2\alpha - 1}}\label{eqn: scalar PS model}
\eeq
The speed of sound determined from (\ref{eqn: Lagrangian}) and (\ref{eq:sound_speed_def}) is
\beq
c_{_S}^{2} = \f{1}{2\,\alpha - 1}.
\label{eqn: sound speed model}
\eeq
We therefore find that the sound speed is a constant. The focus of this paper will be
 on $\alpha > 1$
 for which $c_{_s} < 1$ (\ie~  $c_{_s} < c$ since we work with units $c \equiv 1$).

\section{Inflationary models}
\label{sec:inflation}

\subsection{Chaotic inflation}

Chaotic inflation is usually associated with power law potentials
\beq
V(\phi) = V_{_{0}}\,\phi^{n}~, ~~{\rm where} ~~ V_{_{0}}, ~n ~> ~ 0~.
\label{eq:chaotic}
\eeq
In what follows we shall obtain an expression for $\phi(N)$, with $N$ being the number of
inflationary e-folds
to the end of inflation.
Inflation ends when slow-roll parameters grow and approach the value of unity.
Substituting $\varepsilon_{_V} = 1$ in (\ref{eqn: potential SR1}) we obtain the following
expression for the value of the scalar field when inflation ends
\beq
\frac{\phi_{_{e}}}{\mpl} = \l\{\l(\frac{\mu^{4(\alpha - 1)}}{\alpha}\r)
\l(\frac{3\,\mpl^{4 - n}}{V_{_{0}}}\r)^{\alpha -1}
\l(\frac{n}{\sqrt{2}}\r)^{2\alpha}\r\}^{\frac{1}{\gamma\,(2\alpha - 1)}},
\label{eqn: phi end}
\eeq
where
\beq
\gamma \equiv \frac{2\alpha + n\,\l(\alpha - 1\r)}{2\alpha - 1}, ~~ \mu \equiv \frac{M}{\mpl}~.
\label{eqn: gamma}
\eeq

The number of e-folds to the end of inflation is
\beq
N = -\int^{\phi}_{\phi_{_{e}}}\l(\frac{H}{\dot{\phi}}\r)\d\,\phi.
\label{eqn: efold}
\eeq
Substituting for ${\dot \phi}$ from (\ref{eqn: phidot})
 and for $\phi_{_{e}}$ from (\ref{eqn: phi end})
we obtain the following simple expression
\beq
\frac{\phi(N)}{\mpl} = C_{_{1}}^{1/\gamma}\l(N\gamma + \frac{n}{2}\r)^{\frac{1}{\gamma}}~,
\label{eqn: phi N}
\eeq
where
\beq
  C_{_{1}} = \l\{\l(\frac{n\,\mu^{4(\alpha - 1)}}{\alpha}\r)
\l(\frac{6\,\mpl^{4 - n}}{V_{_{0}}}\r)^{\alpha -1}
\r\}^{\frac{1}{2\alpha - 1}}~,
\label{eqn: C1}
\eeq
which reduces to the standard result
\beq
\frac{\phi(N)}{\mpl} = \sqrt{n\l(2\,N + \frac{n}{2}\r)}
\hspace{0.5cm}  \mathrm{when} \hspace{0.5cm} \alpha\,=\,1~.
\eeq

We now use the results of the preceding section to determine spectral
indices for scalar and tensor perturbations. Substituting (\ref{eq:chaotic}) in
(\ref{eqn: scalar PS model}) we find
\beq
\mathcal{P}_{_{S}}(k) = A_{_{S}}\l(\frac{\phi}{\mpl}\r)^{\gamma + n}_{aH\, =\, c_{_s}k},
\label{eqn: PS phi-n potential}
\eeq
where $\gamma$ was defined in (\ref{eqn: gamma}) and the amplitude $A_{_{S}}$ is given by
\beq
A_{_{S}} = \l(\frac{1}{72\pi^{2}c_{_S}}\r)\l\{\l(\frac{\alpha\, 6^{\alpha}}{n^{2\alpha}\,\mu^{4(\alpha-1)}}\r)
\l(\frac{V_{_{0}}}{\mpl^{4 - n}}\r)^{3\alpha - 2}
\r\}^{\frac{1}{2\alpha - 1}}.
\label{eqn: As}
\eeq

The scalar spectral index $n_{_{S}}$ is defined as
\beq
n_{_{S}} - 1 \equiv  \frac{\d\, \mathrm{ln} \mathcal{P}_{_{S}}}{\d\, \mathrm{ln} k}.
\label{eqn: ns definition}
\eeq
Since $H \simeq$ constant during slow roll inflation and $c_{_S}$ is constant for our model,
it turns out that at sound horizon exit ($a\,H = c_{_S}k$)
\beq
\frac{\d\, }{\d\, \mathrm{ln} k}\, \simeq\,  -   \frac{\d\,}{\d\, N} ~,\label{eqn: dlnk dN}
\eeq
where $N$ is the number of e-folds counted from the end of inflation.
Therefore, from
(\ref{eqn: PS phi-n potential}), (\ref{eqn: ns definition}), (\ref{eqn: dlnk dN}) we have
\beq
n_{_{S}} - 1 = -(\gamma + n)\l(\f{1}{\phi}\frac{\d\, \phi}{\d\, N}\r).
\eeq
Substituting for $\phi(N)$ from (\ref{eqn: phi N}) into the above equation gives
the elegant and simple result
\beq
n_{_{S}}= 1 -2\l(\f{\gamma + n}{2N\gamma + n}\r)
\label{eqn: ns phi n potential}
\eeq
with $n$ defined in (\ref{eq:chaotic}) and $\gamma$ defined in (\ref{eqn: gamma}).
The running of the spectral index is given by
\beq
\frac{\d n_{_{S}}}{\d\, \mathrm{ln} k} = -\frac{1}{1+n/\gamma}(n_{_{S}} - 1)^2~.
\label{eq:running}
\eeq
For $\alpha = 1$, eqn (\ref{eqn: ns phi n potential}) reduces to the standard
result for large field inflationary models with
a canonical kinetic term, namely
\beq
n_{_{S}}= 1 -\f{2(n + 2)}{4N + n}~.
\eeq

\begin{figure}[t]
\scalebox{0.89}[1]{\includegraphics{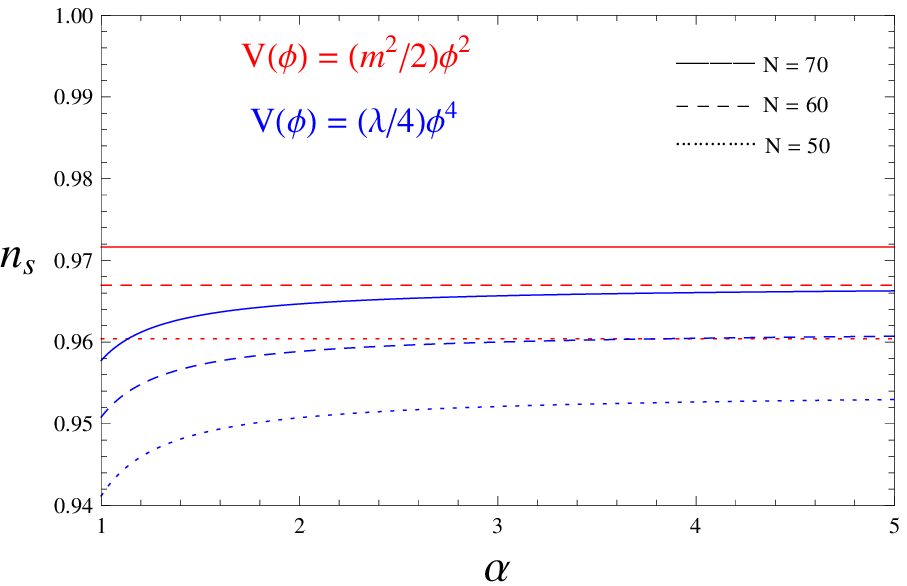}}
\scalebox{0.89}[1]{\includegraphics{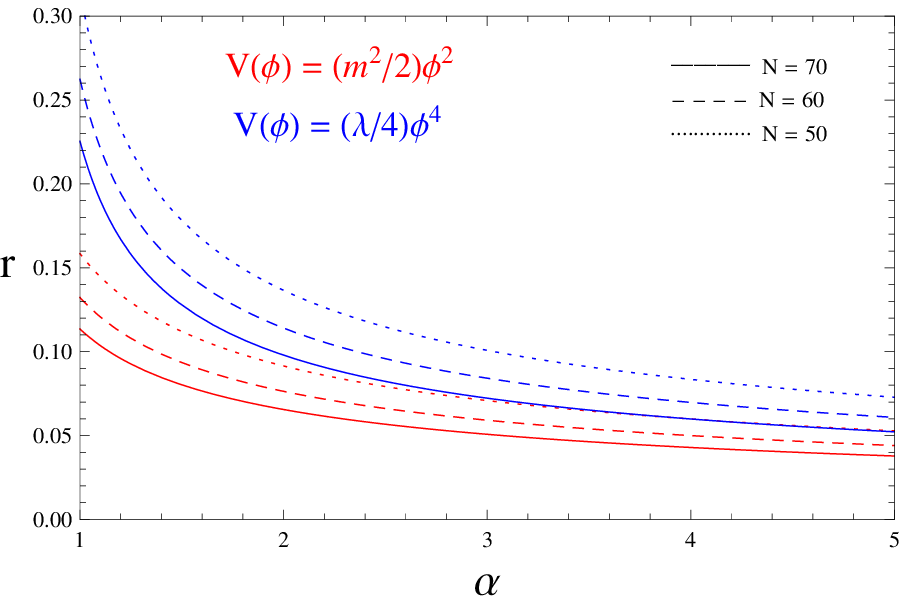}}
\caption{
The scalar spectral index $n_{_{S}}$ (left panel)
and the tensor-to-scalar ratio $r$ (right panel) are shown as functions of $\alpha$ defined in
(\ref{eqn: Lagrangian}).
Red curves corresponds to $V(\phi) = (m^{2}/2)\phi^{2}$ while blue curves represent $V(\phi) = (\lambda/4)\phi^{4}$.
Increasing $\alpha$ leads to a decrease in $r$ for $m^{2}\phi^{2}$ and to a simultaneous
decrease in $r$ and increase in $n_{_{S}}$ for $\lambda\phi^{4}$.
As we shall see in figure \ref{fig:ns-r chaotic}
 this allows the $\lambda\phi^{4}$
potential to agree much better with CMB data than in the canonical case.
}
\label{fig:ns}
\end{figure}

Several points need to be noted here:

\begin{itemize}
  \item Substituting $n=2$ in
(\ref{eqn: ns phi n potential}), which corresponds to  $V(\phi) = \frac{1}{2}m^{2}\phi^{2}$,
we obtain
\beq
   n_{_{S}}= 1 -\l(\f{4}{2N + 1}\r).
\eeq
Surprisingly this result {\em does not} depend upon the value of $\alpha$ and so we conclude
that
the scalar spectral index  $n_{_{S}}$ for this
potential is identical for canonical and non-canonical Lagrangians of the form (\ref{eqn: Lagrangian}) !
%{\em independent of the value of the parameter $\alpha$ in the

\item Substituting $n=4$ in
(\ref{eqn: ns phi n potential}), which corresponds to  $V(\phi) = \frac{1}{4}\lambda\phi^{4}$,
we obtain
\beq
n_{_{S}}= 1 -\l(\f{\gamma + 4}{N\gamma + 2}\r).
\eeq
Since $\gamma$ in (\ref{eqn: gamma}) varies from $\gamma = 2$ ($\alpha = 1$) to
$\gamma = 3$ ($\alpha  \gg 1$) we find that
the scalar spectral index $n_{_{S}}$ (for $N = 60$) {\em increases}
from $n_{_{S}} = 0.951$ ($\alpha = 1$) to $n_{_{S}} = 0.962$ ($\alpha  \gg 1$);
see figure \ref{fig:ns}.

\item Generically,
for $n>0$ in  $V(\phi) = V_{_{0}}\,\phi^{n}$,
 the value of  $n_{_{S}}$ asymptotically approaches the constant value
\beq
   n_{_{S}}= 1 -\f{3n + 2}{N(n + 2) + n} \hspace{0.5cm}  \mathrm{when} \hspace{0.5cm} \alpha\,\gg\, 1.
\eeq
 Indeed, from the left panel of figure \ref{fig:ns} we see that as
$\alpha$ increases the value of $n_{_{S}}$ for the $\lambda\phi^{4}$ potential with
$N=60 (70)$ approaches the value of $n_{_{S}}$ for the $m^2\phi^2$ potential with the lower
value of $N=50 (60)$. %The right panel of figure \ref{fig:ns} shows that,
%for $\alpha \ggeq 3$, the value of $r$ for the $\lambda\phi^{4}$ potential with $N=70$
%virtually coincides with the value of $r$ for the $m^2\phi^2$ potential with $N=50$.

\end{itemize}

Turning now to the tensor power spectrum and substituting $V(\phi) = V_{_{0}}\,\phi^{n}$
in  (\ref{eqn: tensor PS}) we get
\beq
\mathcal{P}_{_{T}}(k) = \l(\frac{2\,}{3\,\pi^{2}}\r)\l(\f{V_{_{0}}}{\mpl^{4 - n}}\r)\l(\f{\phi}{\mpl}\r)^{n}_{aH\, =\, k}.
\label{eqn: tensor PS pni-n pot}
\eeq
An analytical form for the tensor spectral index
\beq
n_{_{T}} \equiv  \frac{\d\, \mathrm{ln} \mathcal{P}_{_{T}}}{\d\, \mathrm{ln} k} ~,
\label{eqn: nt definition}
\eeq
can be obtained by substituting $\phi(N)$ from (\ref{eqn: phi N}) into
 (\ref{eqn: tensor PS pni-n pot})
and using (\ref{eqn: dlnk dN}), it follows that
\beq
n_{_{T}} = -\f{2\,n}{2\,N\gamma + n}~,
\label{eqn: nt phi n potential}
\eeq
where $\gamma$ was defined in (\ref{eqn: gamma}).

We now proceed to obtain the tensor-to-scalar ratio
\beq
r \equiv \f{\mathcal{P}_{_{T}}}{\mathcal{P}_{_{S}}}.
\label{eqn: T-to-S def}
\eeq
When evaluating $r$ it is important to keep in mind that
the expression (\ref{eqn: PS phi-n potential}) for  the scalar power spectrum $\mathcal{P}_{_{S}}(k)$
is evaluated at sound horizon exit ($aH\, =\, c_{_S}k$) while the
corresponding
 expression (\ref{eqn: tensor PS pni-n pot}) for the tensor power spectrum is evaluated at
horizon exit ($aH\, =\,k$).
However, since $H\,\simeq$ constant during slow roll, and the speed of sound $c_{_s}$
 does not depend upon time,
the value of the field $\phi$ at
sound horizon exit is approximately the same as at horizon exit \cite{Garriga-1999}.
Therefore, substituting  (\ref{eqn: PS phi-n potential}) and (\ref{eqn: tensor PS pni-n pot}) in
(\ref{eqn: T-to-S def}) and using $\phi(N)$ from Eq.(\ref{eqn: phi N}) we finally get
\beq
r =  \l(\f{1}{\sqrt{2\,\alpha - 1}}\r)\l(\f{16\, n}{2\,N\gamma + n}\r).
\label{eqn: T-to-S phi-n pot}
\eeq
From (\ref{eqn: T-to-S phi-n pot}) and (\ref{eqn: gamma}) we find
$$r =  \f{16\, n}{4\,N + n} \hspace{0.5cm}  \mathrm{when} \hspace{0.5cm} \alpha\,=\,1$$
which is the standard value for canonical scalars.

The prefactor $(2\,\alpha - 1)^{-1/2}$ in (\ref{eqn: T-to-S phi-n pot}) informs us
that the value of $r$ {\em decreases} as $\alpha$ is increased.
In other words, non-canonical models with $\alpha > 1$ generically give rise to a lower
tensor-to-scalar ratio than canonical models, which is one of the central results of this paper
and is illustrated in figure \ref{fig:ns}.
A lower value of $r$ helps in making the $\lambda\phi^4$ potential come into agreement with
CMB data, as shown later in this section.
Indeed, the right panel of fig. \ref{fig:ns}  illustrates that,
for $\alpha \ggeq 3$, the value of $r$ for the $\lambda\phi^{4}$ potential with $N=70$
overlaps with the value of $r$ for the $m^2\phi^2$ potential with $N=50$.

Using (\ref{eqn: T-to-S phi-n pot}) and (\ref{eqn: nt phi n potential}) one finds the following
consistency relation for slow roll inflation with a non-canonical scalar field \cite{Garriga-1999}
\beq
r = -8\,c_{_S}\,n_{_{T}}.
\label{eqn: consistency reln}
\eeq
Note that the presence of the sound speed $c_{_S}$ in (\ref{eqn: consistency reln})
causes the consistency relation to differ from the
canonical case for which $r = -8\,n_{_{T}}$.
Substituting for $c_{_S}$ from (\ref{eqn: sound speed model}) we find
\beq
r = -\frac{8\,n_{_{T}}}{\sqrt{2\alpha - 1}}~.
\label{eq:consistency}
\eeq
Since $\alpha > 1$ we find that the value of $r$
for non-canonical models
is generically {\em smaller} than that for canonical models with identical
values of $n_{_{T}}$.

For comparison note that a prominent example of a non-canonical scalar is
provided by the tachyon model \cite{Fairbairn-2002,Steer-2004,Campo-2009} for which
${\cal L}(X,\phi) = -V(\phi)\sqrt{1 - 2\,X}$ and
the speed of sound and
  equation of state are related through $c_{_S}^{2} = -\,w$.
Since $w \simeq -1$ during slow roll, it follows that $c_{_S} \simeq 1$
for this class of models. Consequently, the consistency relation for slow roll inflation in
tachyon models is the same as that for canonical scalar fields,
namely $r = -8\,n_{_{T}}$. This is certainly not the case for our model (\ref{eqn: Lagrangian})
since {\em even in the slow roll regime} an appropriate choice of the free parameter
$\alpha$ can modify the speed of sound and hence the consistency relation
(\ref{eqn: consistency reln}).
We therefore conclude that the consistency relation
can help differentiate between our non-canonical model (\ref{eqn: Lagrangian}) and standard
(slow roll) inflationary models as well as tachyon models.
Eqn. (\ref{eq:consistency}) therefore emerges as a {\em smoking gun} test
for the inflationary models examined in this paper \cite{Garriga-1999}.

\bigskip
\leftline{\underline{\bf CMB normalization}}

\bigskip

The scalar power spectrum (\ref{eqn: PS phi-n potential}) requires to be normalized using
CMB observations. This can be done in a straightforward manner by noting that
$\mathcal{P}_{_{S}}(k_{\ast}) \simeq 2.4\,\times\, 10^{-9}$, where $k_{\ast} = 0.002\, \mathrm{Mpc}^{-1}$ is the pivot scale \cite{Komatsu-2011}.
Substituting the solution $\phi(N)$ from (\ref{eqn: phi N}) into
 the expression (\ref{eqn: PS phi-n potential})  and rearranging, we get for the potential
$V = V_0\phi^n$
\beq
\frac{V_{_0}}{\mpl^{4-n}} =
\l\{
\f{12\pi^{2}\,n\,\mathcal{P}_{_S}(k_{\ast})}{\sqrt{2\,\alpha -1}}
\l(\frac{\alpha}{n}\l(\frac{1}{6\mu^{4}}\r)^{\alpha-1}\r)^{\frac{n}{\gamma(2\alpha - 1)}}
\l(\f{2}{2\,N_{\ast}\gamma + n}\r)^{\frac{\gamma + n}{\gamma}}
\r\}^{\frac{\gamma(2\alpha - 1)}{2\,\alpha}},
\label{eqn: v-0}
\eeq
where $\gamma$ was defined in (\ref{eqn: gamma})
and $N_{\ast}$ denotes the number of e-folds from the end of inflation to the pivot scale.
%\,\simeq\,60$ e-folds for the pivot scale.
%The above equation allows one to determine the value of $V_0$ in the potential $V = V_0\phi^n$.
The above equation allows one to determine
the value of the mass in $V(\phi) = m^{2}\phi^{2}/2$ and that
of $\lambda$ in $V(\phi) = \lambda\phi^{4}/4$.

\begin{figure}[t]
\begin{center}
%\scalebox{0.89}[1.05]{\includegraphics{mass-m.eps}}
\scalebox{1.2}[1.2]{\includegraphics{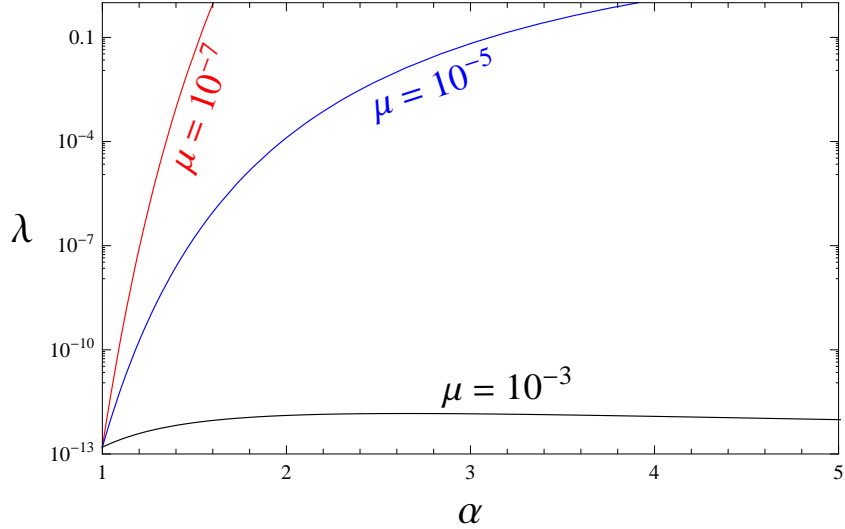}}
\caption{The $V = \frac{1}{4}\lambda\phi^4$ inflationary model is tested against observations giving
the above CMB normalized value of $\lambda$ which is shown plotted as a function of $\alpha$.
Three different values of $\mu\, \equiv\, M/\mpl$ are shown. Note that for lower values
$M \lleq 10^{14}$ GeV the value of $\lambda$ can be as large as $\lambda \sim O(1)$ for
$\alpha \sim {\rm few}$. $M$ and $\alpha$ are defined in (\ref{eqn: Lagrangian}).
 The perturbation spectrum for this model provides a decent fit
to CMB data, as demonstrated in figure \ref{fig:ns-r chaotic}.
%In this figure, the CMB normalized value of $\lambda$  is plotted as a function of $\alpha$ for three different values of $\mu\, \equiv\, M/\mpl$.
}
\label{fig:lambda}
\end{center}
\end{figure}

\begin{itemize}

\item
Substituting $n=2$ and $V_{_0} = m^2/2$ into (\ref{eqn: v-0}) we obtain
for the potential $V(\phi) = m^{2}\phi^{2}/2$
\beq
\frac{m}{\mpl} = \l(\frac{m_{_1}}{\mpl}\r)^{\f{2\,\alpha -1}{\alpha}}
\l\{\l(\f{1}{2\,\alpha - 1}\r)^{\f{2\,\alpha -1}{2}}\l(\f{\alpha}{\l(3\mu^{4}\r)^{\alpha - 1}}\r)
\r\}^{\frac{1}{2\alpha}},
\label{eqn: mass m}
\eeq
where $\mu\, \equiv\, M/\mpl$ and
\beq
\frac{m_{_1}}{\mpl} = \sqrt{\frac{24\,\pi^{2}\,\mathcal{P}_{_{S}}(k_{\ast})}{(2\,N_{\ast} + 1)^{2}}}\; %\simeq\; 6.2\,\times\, 10^{-6}\nn
\label{eq:mass0}
\eeq
is the CMB normalized mass associated with canonical inflation, which corresponds to
$\alpha=1$ in
(\ref{eqn: Lagrangian}). %In other words,
%$m = m_{_1}$ in (\ref{eqn: mass m}) for
%canonical values $\mu=1$ and $\alpha=1$.
Setting $N_{\ast} = 60$ in (\ref{eq:mass0}) results in the standard value
$\frac{m_{_1}}{\mpl} \simeq\; 6.2\,\times\, 10^{-6}$.

\item
Substituting $n=4$ and $V_{_0} = \lambda/2$ into (\ref{eqn: v-0}) we obtain
for the potential $V(\phi) = \lambda\phi^{4}/4$
\beq
 \lambda = 4\l\{
 \f{32 \lambda_{_1}(N_{\ast} + 1)^{3}}{\sqrt{2\,\alpha -1}}
 \l(\frac{\alpha}{4}\l(\frac{1}{6\mu^{4}}\r)^{\alpha-1}\r)^{\frac{2}{3\alpha - 2}}
 \l(\f{1}{N_{\ast}\gamma + 2}\r)^{\frac{\gamma + 4}{\gamma}}
 \r\}^{\frac{3\,\alpha - 2}{\alpha}},
\label{eqn: lambda}
\eeq
where
\beq
\lambda_{_1}\, =\, \frac{3\,\pi^{2}\,\mathcal{P}_{_{S}}(k^{\ast})}{2(\,N_{\ast} + 1)^{3}}
%\; \simeq\; 1.5\,\times\, 10^{-13},\nn
\label{eq:lambda0}
\eeq
is the CMB normalized dimensionless coupling $\lambda$ associated with canonical inflation.
Setting $N_{\ast} = 60$ in (\ref{eq:lambda0}) results in the standard value
$\lambda_{_1} = 1.5\,\times\, 10^{-13}$.
(Note that $\lambda = \lambda_{_1}$ in (\ref{eqn: lambda}) for
canonical values $\mu=1$ and $\alpha=1$.)
\end{itemize}

\begin{figure}[t]
\scalebox{0.89}[1]{\includegraphics{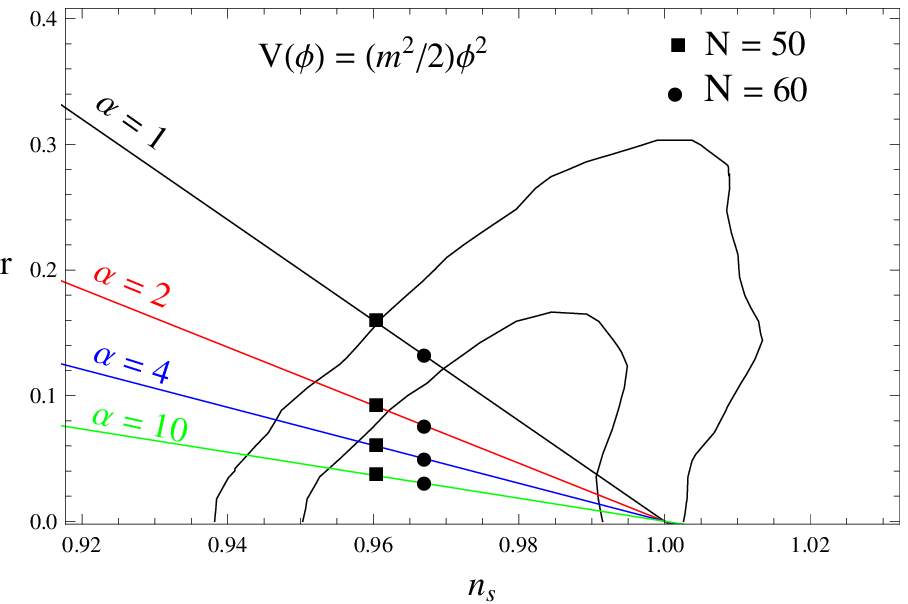}}
\scalebox{0.89}[1]{\includegraphics{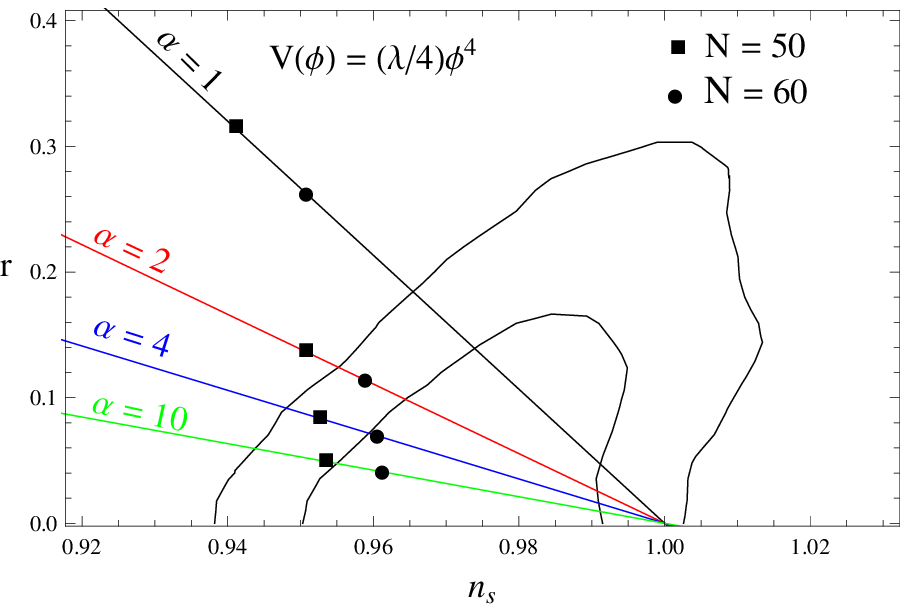}}
\caption{The spectral index $n_{_S}$ and the tensor to scalar ratio $r$
are shown for different values of the parameter $\alpha$ in (\ref{eqn: Lagrangian}),
for chaotic inflation sourced by the $m^2\phi^2$ potential (left) and the $\lambda\phi^4$
potential (right).
The inner and outer contours correspond to $1\sigma$ and $2\sigma$ confidence limits obtained
using WMAP7, BAO and HST data.
$\alpha = 1$ corresponds to canonical scalars which are ruled out
for the $\lambda\phi^4$ model.
Increasing $\alpha$ leads to an increase in $n_{_S}$ and a decrease in $r$ resulting in
a marked improvement of fit for the $\lambda\phi^4$ model.
 $N$ denotes the number of e-folds to
the pivot scale at $k = 0.002\, \mathrm{Mpc}^{-1}$. %Larger values of both $N$ and $\alpha$ appear to be preferred by observations.
}
\label{fig:ns-r chaotic}
\end{figure}

Our results for $\lambda$ are illustrated in figure \ref{fig:lambda}
which contains the following interesting
information. For $\mu < 10^{-3}$ the value of $\lambda$ grows as $\alpha$ increases.
Indeed the growth of $\lambda$ becomes quite spectacular for smaller values of $\mu$.
For instance when $\mu = 10^{-5}$,
 $\lambda$ grows by over 10 orders of magnitude from
its canonical value of $\lambda \sim 10^{-13}$ for $\alpha=1$, to
$\lambda \simeq 1$ for $\alpha = 4$.
Furthermore figure \ref{fig:ns} informs us
 that for  $\alpha = 4$ the tensor-to-scalar ratio in this model is $r \sim\,0.1$,
 which is in good agreement with observations -- see fig. \ref{fig:ns-r chaotic}.
One might also like to note that smaller values of $\mu\, \equiv\, M/\mpl$ are more
physically appealing since they correspond to sub-Planckian mass scales, with
$\mu \sim 10^{-3} \Rightarrow M \sim 10^{16}$ GeV being the energy scale of Inflation.
[$M$ and $\alpha$ have been defined in (\ref{eqn: Lagrangian}) for our model.]
We therefore come to the conclusion that $\lambda\phi^4$ Inflation, which runs into CMB trouble in
the canonical case,
reverts back into favor when viewed within a non-canonical setting.
(Similar results have been recently obtained in \cite{shinji} for an inflationary model
with a field derivative coupling to gravity; see also \cite{jamia}.)

\subsection{The exponential potential}
\label{sec:expo}

Another important example of a `large field' potential is the exponential
\beq
V(\phi) = V_{_0}\exp\l(-\sqrt{\frac{2}{q}}\,\frac{\phi}{\mpl}\r)~.
\label{eqn: exp pot}
\eeq
A spatially flat universe dominated by a canonical scalar field with this potential expands as
 a power law $a(t) \propto t^{q}$ \cite{PLI}.
It is easy to see that in this case
the slow roll parameters in (\ref{eqn: slow roll}) \& (\ref{eqn: potential SR1})
 are constants
$\varepsilon = \varepsilon_{_V} = 1/q$, making a natural exit from inflation impossible.
Remarkably this is not the case for non-canonical models for which the slow roll parameter
in (\ref{eqn: potential SR1}) acquires the form
\beq
\varepsilon_{_V} = \l[\frac{1}{\alpha q^\alpha}
\l(\frac{3\,M^{4}}{V(\phi)}\r)^{\alpha -1}\r]^{\frac{1}{2\alpha -1}}
\label{eqn: SR exp pot}
\eeq
which reduces to $\varepsilon_{_V} = 1/q$ for $\alpha = 1$. Clearly
for $\alpha > 1$
the value of $\varepsilon_{_V}$ can be extremely small if $V \gg M^4$. Thus the slow roll parameter
evolves from $\varepsilon_{_V} \ll 1$ initially, to $\varepsilon_{_V} \sim 1$ as $\phi$ rolls down
its potential and $M^4/V$ increases
\footnote{As mentioned earlier, we only consider the case where the parameter $\alpha > 1$.
This will ensure that the speed of sound $c_{_s}$ is less than the speed of light.
It is only when $\alpha > 1$ $\varepsilon_{_V}$ evolves from
$\varepsilon_{_V} \ll 1$ and crosses $\varepsilon = 1$ when scalar field rolls down the potential.
However, when $\alpha < 1$, the slow roll parameter $\varepsilon_{_V}$ in fact decreases as the scalar field rolls down the potential which makes exit from inflation even more difficult to attain.}.
Inflation based on the exponential potential with a non-canonical scalar therefore
has a graceful exit !

From (\ref{eqn: phidot}) and (\ref{eqn: SR exp pot}) one can show that in the slow roll regime
\beq
\varepsilon_{_V} = \frac{1}{\sqrt{2\,q}}\l(\frac{1}{\mpl}\r)\l(\frac{\dot{\phi}}{H}\r) = -\frac{1}{\sqrt{2\,q}}\l(\frac{1}{\mpl}\r)\l(\frac{\d \phi}{\d N}\r)~.
\label{eqn: SR exp eq1}
\eeq
Differentiating (\ref{eqn: SR exp pot}) with respect to $N$ and using the above equation,
we arrive at the following equation of motion for $\varepsilon_{_V}$:
\beq
\frac{\d \varepsilon_{_V}}{\d N} = -2\l(\frac{\alpha - 1}{2\alpha -1}\r)\varepsilon_{_V}^{2}
\label{eqn: SR exp EOM}
\eeq
which can be solved to give
\beq
\varepsilon_{_V} = \frac{2\alpha -1}{2\alpha -1\, +\, 2(\alpha -1)N} ~.
\label{eqn: SR exp soln-1}
\eeq
Equation (\ref{eqn: SR exp EOM}) ensures that when $\alpha = 1$, $\varepsilon_{_V}$ is identically constant and to be
consistent with the standard result, its value should be $q^{-1}$.
However, from the solution (\ref{eqn: SR exp soln-1}), it turns out that for $\alpha \neq 1$, the slow roll parameter $\varepsilon_{_V}$ is independent of the value of $q$ and astonishingly
the solution does not converge to the standard result
({\it viz.~$\varepsilon_{_V} = 1/q$}) when $\alpha$ is set to unity.
As we shall see shortly, in addition to $\varepsilon_{_V}$
 neither $n_{_{S}}$ nor $r$ depend on $q$, and the
reason for this puzzling behavior
will be explained at the end of this section.

\begin{figure}[t]
\begin{center}
\scalebox{1.2}[1.2]{\includegraphics{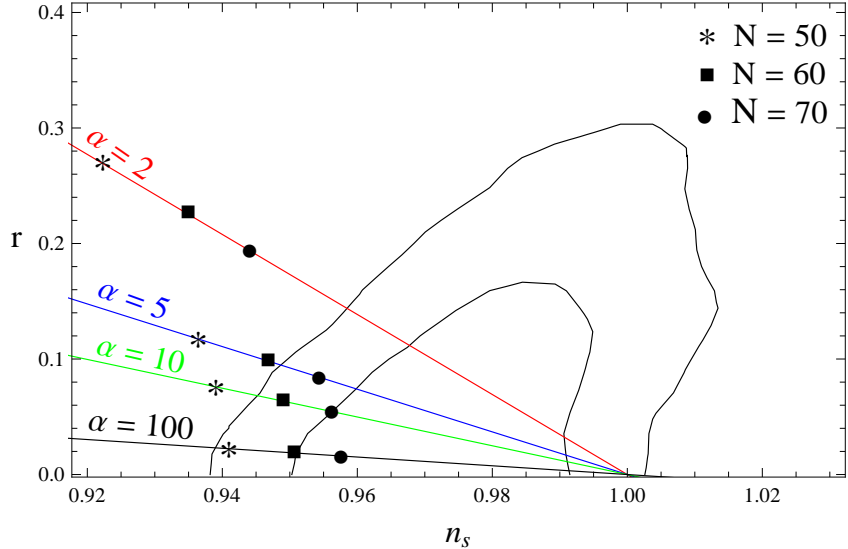}}
\caption{The spectral index $n_{_S}$ and the tensor to scalar ratio $r$
are shown for different values of the parameter $\alpha$ in (\ref{eqn: Lagrangian}),
for inflation sourced by the exponential potential (\ref{eqn: exp pot}).
 As earlier, $N$ denotes the number of e-folds to
the pivot scale at $k = 0.002\, \mathrm{Mpc}^{-1}$,
and inner and outer contours correspond to $1\sigma$ and $2\sigma$ confidence limits obtained
using WMAP7, BAO and HST data.
Increasing $\alpha$ leads to an increase in $n_{_S}$ and a decrease in $r$.
 Larger values of $\alpha$ therefore appear to be preferred by observations.
}
\label{fig:ns-r-exp-pot}
\end{center}
\end{figure}

For the exponential potential, the expression for scalar power spectrum (\ref{eqn: scalar PS model}) reduces to
\beq
\mathcal{P}_{_{S}}(k) = A_{_{S}}\exp\l[-\l(\frac{3\alpha - 2}{2\alpha -1}\r)\sqrt{\frac{2}{q}}\,\frac{\phi}{\mpl}\r],
\label{eqn: PS exp pot}
\eeq
where the amplitude $A_{_{S}}$ is given by
\beq
A_{_{S}} = \l(\frac{1}{72\pi^{2}c_{_S}}\r)\l\{\l(\frac{\alpha\, (3\,q)^{\alpha}}{\mu^{4(\alpha-1)}}\r)
\l(\frac{V_{_{0}}}{\mpl^{4}}\r)^{3\alpha - 2}
\r\}^{\frac{1}{2\alpha - 1}}.
\label{eqn: As exp pot}
\eeq
Using Eqs.(\ref{eqn: dlnk dN}) and (\ref{eqn: SR exp eq1}), the scalar spectral index (\ref{eqn: ns definition}) for the exponential potential turns out to be
\beq
1-n_{_{S}} = 2\,\l(\frac{3\alpha -1}{2\alpha -1}\r)\varepsilon_{_V}~.
\eeq
Substituting $\varepsilon_{_V}$ from (\ref{eqn: SR exp soln-1}) we get
\beq
1 - n_{_{S}} =  2\,\l(\frac{3\alpha -1 }{2\alpha -1\, +\,
2\,N\,(\alpha -1) }\r) ~, ~~ \alpha > 1~.
\label{eqn: ns exp pot}
\eeq
Similarly, using Eqs.(\ref{eq:tensor}), (\ref{eqn: T-to-S def}),
(\ref{eqn: SR exp pot}),(\ref{eqn: PS exp pot})
and (\ref{eqn: SR exp soln-1}),
we get the following result for the tensor to scalar ratio
\beq
r = \frac{16\, \sqrt{2\alpha -1}\, }{2\alpha -1\,
+\, 2\,N\,(\alpha -1) } ~,  ~~\alpha > 1~.
\label{eqn: r exp pot}
\eeq
%\item From the above expression for $n_{_S}$ and $r$ it turns out that
%for $\alpha = 2$, $n_{_S} = 0.935$ and $r = 0.225$ when $N = 60$ whereas for $N = 70$ their values are $0.944$ and $0.194$, respectively.
% These are clearly outside the $95\%$ CL limit ... see Fig.\ref{fig:ns-r-exp-pot}
% However, for large value of $\alpha$, say for $\alpha > 5$, the values $n_{_S}$ and $r$ are with the observational bonds as indicated by Fig.\ref{fig:ns-r-exp-pot}.

Our results for $n_{_S}$ and $r$ are shown together with CMB constraints in figure \ref{fig:ns-r-exp-pot}. We find that for $\alpha > 5$ the exponential potential can be accommodated by
observations.

As mentioned earlier, the expressions for $n_{_{S}}$ and $r$ do not depend on the
value of $q$ in (\ref{eqn: exp pot}) when $\alpha \neq 1$.
The reason for this stems from the fact that under the transformation $\phi \rightarrow \psi \equiv (\sqrt{2/q}\,)\phi$, the Lagrangian (\ref{eqn: Lagrangian}) with an exponential potential becomes
\beq
{\cal L}(\widetilde{X},\psi) = \widetilde{X}\l(\frac{\widetilde{X}}{\widetilde{M}^{4}}\r)^{\alpha-1} -\; V_{_0}\exp\l(-\,\frac{\psi}{\mpl}\r),
\eeq
where
\beq
\widetilde{X} = \frac{1}{2}\pa_{\mu}\psi\; \pa^{\mu}\psi
\hspace{0.5cm} \mathrm{and} \hspace{0.5cm}
\widetilde{M} = M\,\l(\frac{2}{q}\r)^{\frac{\alpha}{4(\alpha -1)}}.
\eeq
We therefore find that it is possible to absorb the parameter $q$  into
a redefinition of the mass $M$ without altering the structure of the kinetic term in the Lagrangian.
(This is possible only for $\alpha \neq 1$.)
     As the  mass $M$ in the kinetic term does not explicitly influence the value of
either $n_{_{S}}$ or $r$, it is but natural to expect neither of these quantities
 to dependent on the parameter $q$ in the potential (\ref{eqn: exp pot}).
  %(One might note that the parameter $q$ can be absorbed into a `renormalization' of $M$
%without altering the structure of the kinetic term in the Lagrangian
%only for $\alpha \neq 1$.)

\section{Scalar field oscillations and the equation of state}
\label{sec:motion}

Non-canonical scalar field models have difficulty in oscillating which could make
reheating problematic in this scenario. This can easily be seen from the following
argument. Conventionally, after inflation has ended, the inflaton field commences
to oscillate about the minimum of its potential. During any given oscillation the field
amplitude is bounded by points where the field trajectory reverses, so that ${\dot\phi} = 0$.
This usually occurs in regions where $V(\phi) \neq 0$ and $V'(\phi) \neq 0$.
%Conventionally, within the frame work of standard canonical inflation,
%once inflation ends, the scalar field starts oscillating about the minima of the potential.
%During such oscillations, after crossing the minima of the potential,
%the scalar field climbs up the potential to the point $\phi_{_m}$ where $\dot{\phi} = 0$
% with $V'(\phi_{_m}) \neq 0$ and then the field moves back towards the minima of the potential.
 Such oscillations are could be problematic within the non-canonical framework
 since, according to the field equation (\ref{eqn: EOM-model}),
 $|\ddot{\phi}| \to \infty$ when $\dot{\phi} = 0$ and $V'(\phi) \neq 0$.

To resolve this issue, one needs to regularize the field equation so
 that the value of $\ddot{\phi}$ remains finite
 even when $\dot{\phi} \rightarrow 0$.
 With this in mind, we propose a modified version of our model replacing (\ref{eqn: Lagrangian}) by the
following `regularized' Lagrangian:
 \beq
{\cal L}_{_R}(X,\phi) = \l(\frac{X}{1 + \beta}\r)\l(1 + \beta\l(\frac{X}{M^{4}}\r)^{\alpha-1}\r) -\; V(\phi),
\label{eqn: Lagrangian-R}
\eeq
where $\beta$ is a dimensionless parameter.
The above prescription is tantamount
to the addition of a canonical kinetic term to the original Lagrangian in (\ref{eqn: Lagrangian}).
From (\ref{eqn: Lagrangian-R}) one finds that
\begin{itemize}
  \item In the limit when $\beta \rightarrow \infty$, the Lagrangian ${\cal L}_{_R}(X,\phi)$ converges to ${\cal L}(X,\phi)$ defined in (\ref{eqn: Lagrangian}).
      Therefore, all the results obtained in the preceding sections also follow from  ${\cal L}_{_R}(X,\phi)$ in the limit when $\beta \rightarrow \infty$.
  \item When $\beta = 0$ or when $\alpha = 1$,  the Lagrangian ${\cal L}_{_R}(X,\phi)$ reduces to the standard Lagrangian for the canonical scalar field.
\end{itemize}

For the modified model (\ref{eqn: Lagrangian-R}), the equation governing the evolution of the scalar field is given by
\beq
{\ddot \phi} + 3\, H{\dot \phi}\l(\frac{1 + \alpha\,\beta\l(X/M^{4}\r)^{\alpha - 1}}{1 + \alpha(2\alpha -1)\beta\l(X/M^{4}\r)^{\alpha - 1}}\r) + \l(\f{( 1 + \beta)V'(\phi)}{1 + \alpha(2\alpha -1)\beta\l(X/M^{4}\r)^{\alpha - 1}}\r) = 0.
\label{eqn: EOM-model-R1}
\eeq
The above equation reduces to (\ref{eqn: EOM-model}) when $\beta \rightarrow \infty$.
When $\beta = 0$ or $\alpha = 1$, the equation (\ref{eqn: EOM-model-R1}) becomes the usual Klein-Gordon equation for the canonical scalar field.
In the limit when $\beta \gg 1$, the above equation can be approximated as
\beq
{\ddot \phi}\,+\, \f{3\, H{\dot \phi}}{2\alpha -1}\, +\, \l(\f{V'(\phi)}{\epsilon\, +\, \alpha(2\alpha -1)\l(X/M^{4}\r)^{\alpha - 1}}\r)\, =\; 0, ~~ X = \frac{1}{2}{\dot \phi}^2~,
\label{eqn: EOM-model-R}
\eeq
where $\epsilon \equiv (1 + \beta)^{-1}$ is an infinitesimally small correction factor  when $\beta >> 1$.
Equation (\ref{eqn: EOM-model-R}) is the regularized version of the field equation (\ref{eqn: EOM-model}) in which
$\epsilon$ acts as a small correction factor which ensures that $\ddot{\phi}$ remains finite even when
$\dot{\phi} \rightarrow 0$ and $V'(\phi) \neq 0$.
With this correction in place the scalar field can oscillate about the minimum of its
 potential once inflation ends.

We now derive an expression for the average of the
 equation of state parameter for the oscillating scalar field.
For the model (\ref{eqn: Lagrangian-R}), the energy density $\rho_{_{\phi}}$ and pressure $p_{_{\phi}}$ are given by
\ber
\rho_{_{\phi}} &=& \l(\frac{X}{1 + \beta}\r)\l(1 + \beta\,(2\alpha-1)\l(\frac{X}{M^{4}}\r)^{\alpha-1}\r) +\;  V(\phi),\nonumber\\
p_{_{\phi}} &=& \l(\frac{X}{1 + \beta}\r)\l(1 + \beta\l(\frac{X}{M^{4}}\r)^{\alpha-1}\r) -\; V(\phi).
\label{eqn: p-model-R}
\eer
Since $w_{_{\phi}} \equiv p_{_{\phi}}/\rho_{_{\phi}}$, in the limit when $\beta >> 1$ one finds
\beq
1 + \l<w_{_{\phi}}\r> = \l(\frac{2\alpha}{M^{4(\alpha - 1)}}\r)\l< \frac{X^{\alpha}}{\rho_{_{\phi}}}\r>,
\label{eqn: avg w1}
\eeq
where $<\,>$ denotes the average over one oscillation cycle.
As in the case of standard canonical inflation, we shall assume that the time scale of oscillation is much smaller than the time scale of expansion of the universe.
In this limit, the time variation of $\rho_{_{\phi}}$ during any one cycle
is sufficiently small to permit the approximation
$\rho_{_{\phi}} \simeq V(\phi_{_m})$, where $\phi_{_m}$ is the
maximum value of the field during a given cycle.
Using Eq.(\ref{eqn: p-model-R}) and the fact that $\rho_{_{\phi}} \simeq V(\phi_{_m})$, we find that in the limit
$\beta \gg 1$
\beq
\frac{X^{\alpha}}{\rho_{_{\phi}}} = \l(\f{M^{4(\alpha - 1)}}{2\alpha -1}\r)\l(1\, -\, \frac{V(\phi)}{V(\phi_{_m})}\r).
\label{eqn: xalpha avg}
\eeq
From Eqs.(\ref{eqn: avg w1}) and (\ref{eqn: xalpha avg}) it is straight forward to show that
\beq
1 + \l<w_{_{\phi}}\r> = \l(\frac{2\alpha}{2\alpha - 1}\r)
\l(\int^{\phi_{_m}}_{0}\d \phi \l(1\, -\, \frac{V(\phi)}{V(\phi_{_m})}\r)^{\frac{2\alpha - 1}{2\alpha}}\r)
\l[\int^{\phi_{_m}}_{0}\d \phi \l(1\, -\, \frac{V(\phi)}{V(\phi_{_m})}\r)^{-\frac{1}{2\alpha}}\r]^{-1}~.
\label{eqn: avg w}
\eeq

It can be verified that the above expression reduces to the standard result
for the canonical scalar field when the parameter $\alpha$ is set to unity (see Eq.(11) of Ref.\cite{turner-1983}).

\begin{figure}[t]
\begin{center}
\scalebox{1.2}[1.2]{\includegraphics{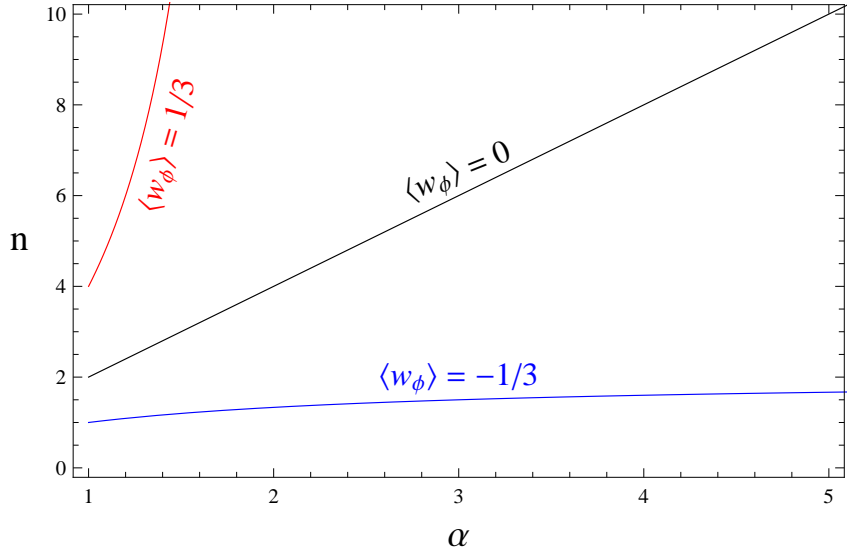}}
\caption{Lines corresponding to $\l<w_{_{\phi}}\r> = 0,\; 1/3$ and $-1/3$
are plotted in the $n\,-\,\alpha$ plane.
}
\label{fig: w-avg}
\end{center}
\end{figure}

For chaotic potentials of the form $V(\phi) = V_{_{0}}\,\phi^{n}$, one integrates (\ref{eqn: avg w})
to determine
\beq
\l<w_{_{\phi}}\r> = \frac{n\, -\, 2\alpha}{n\,(2\alpha -1)\, +\, 2\alpha}~.
\label{eqn: avg w phi n pot}
\eeq
The following points in relation to (\ref{eqn: avg w phi n pot}) deserve special mention:

\begin{enumerate}
  \item For $\alpha = 1$ the above expression reduces to the standard result \cite{turner-1983}:
  \beq
\l<w_{_{\phi}}\r> = \frac{n\, -\, 2}{n\, +\, 2}~.
\eeq
  \item For $n = 2\alpha$, $\l<w_{_{\phi}}\r> = 0$ and therefore
  the oscillating scalar field cosmologically mimics the dynamics of the pressureless matter (dust).
  For example, when $\alpha = 2$, a scalar field oscillating
  about the minimum of a $\lambda\phi^{4}$ potential
  would effectively behave as dust, in stark contrast to the canonical case for which
$\l<w_{_{\phi}}\r> = 1/3$.
  \item Equation (\ref{eqn: avg w phi n pot}) can be rewritten as
  \beq
  n = \frac{2\alpha\,\l(1 + \l<w_{_{\phi}}\r>\r)}{1\, -\,(2\alpha -1)\l<w_{_{\phi}}\r>}.
  \label{eqn: n avg w}
  \eeq
  When $\l<w_{_{\phi}}\r> = 1/3$ the above relation becomes
  \beq
  n = \frac{4\alpha}{2 - \alpha}.
  \eeq
  Since $n > 0$, the above equation informs us that for the oscillating scalar field to behave
as radiation the value of the parameter $\alpha$ in (\ref{eqn: Lagrangian-R}) must be less than $2$.
  \item For $\l<w_{_{\phi}}\r> = - 1/3$, it follows from equation (\ref{eqn: n avg w}) that
  \beq
  n = \frac{2\alpha}{1\, +\, \alpha}.
  \eeq
From the above equation it is clear that asymptotically when  $\alpha \rightarrow \infty$, $n \rightarrow 2$.
Therefore, for the oscillating scalar field to behave as a fluid with $\l<w_{_{\phi}}\r> < - 1/3$, the value of $n$ in the potential $V(\phi) = V_{_{0}}\,\phi^{n}$ must be less than $2$.

These features have been illustrated in Fig.~\ref{fig: w-avg}.
\end{enumerate}

The fact that our new formalism (\ref{eqn: Lagrangian-R}) permits non-canonical fields
to oscillate, and the observation that for sufficiently large values of $\alpha$
the exponential potential provides a plausible inflationary model, allows us to introduce
the following inflationary potential
\beq
V = V_0(\cosh{\lambda\phi} - 1)~,
\label{eq:cosh}
\eeq
which had been suggested as a dark matter candidate in \cite{sw00}.
The potential (\ref{eq:cosh}) has the following asymptotic forms:
\ber
V(\phi) &=& {\tilde{V_0}}\,e^{-\lambda\phi} ~~~~{\rm for} ~~~~|\lambda\phi| \gg 1 ~~~~(\phi < 0)\nonumber\\
V(\phi) &=& {\tilde{V_0}}\,(\lambda\phi)^2 ~~~~{\rm for} ~~~~|\lambda\phi| \ll 1~.
\eer
A non-canonical scalar field rolling down such a potential could initially give rise to
inflation following the discussion in section \ref{sec:expo}. Later, when $\phi$ dropped
to sufficiently small values the scalar field would oscillate, and its mean equation of state
would be given by
\beq
\l<w_{_{\phi}}\r> = \frac{1-\alpha}{3\alpha - 1}
\label{eq:EOS}
\eeq
implying $-1/3 < \l<w_{_{\phi}}\r> < 0$ for $\alpha > 1$. Whether the
standard reheating mechanism will work for such a potential
remains an interesting open question.

\section{Conclusions}
\label{sec:conclusions}
Focussing on a particular class of non-canonical scalar field models, namely
those based on the Lagrangian density (\ref{eqn: Lagrangian}), we have shown that such models provide
an attractive setting in which to re-examine Inflation.
Our non-canonical inflationary
 models generically possess a lower tensor-to-scalar ratio and, in some cases,
also a higher value of the scalar spectral index as compared to their canonical counterparts.
This leads to a better agreement with observations as illustrated by the
$\lambda\phi^4$ potential which agrees with CMB data for values of $\lambda$ as large as
$\lambda \sim O(1)$.

It is well known that the exponential potential, in the canonical case, gives rise to
(eternal) power law inflation and therefore faces a serious `graceful exit' problem.
On examining its non-canonical counterpart we find that inflation for this potential
is no longer eternal, and, like its sister potentials $V \propto \phi^n$, the exponential
exits the slow-roll regime and therefore
does not face a graceful exit problem. Furthermore it turns out that
 the tensor-to-scalar ratio is smaller for the exponential than in the canonical case,
leading to better agreement with observations.

We also find that, under certain conditions, the inverse power law potential and other
steep potentials commonly associated with dark energy  in the canonical case,
can source inflation
for non-canonical fields.
This considerably broadens the class of potentials used for inflationary model
building and could open up the possibility of constructing models of Quintessential-Inflation in
the non-canonical setting.
Whether steep potential will satisfy
the stringent constraints on $\lbrace r, n_S\rbrace$ imposed by the CMB will form
the subject of a future investigation.

Our study in section \ref{sec:motion} has shown that
non-canonical scalar fields have difficulty in oscillating. Since oscillations are an
integral part of preheating scenario's \cite{Bassett-2005,reheat,reheating-review,Allahverdi} this difficulty could prove calamitous
for inflationary models in the non-canonical setting.
%A possible resolution of this problem is provided in the same section; also see \cite{underwood}.
The addition of a canonical kinetic term resolves this issue as shown in
section \ref{sec:motion}; also see \cite{underwood}.

To summarize, we have shown in the context of large field models, that inflation becomes
easier to realize if one generalizes the scalar field Lagrangian to accommodate non-canonical scalars.
(Similar results have been obtained for an affiliated class of models in \cite{gwyn}.)
Our treatment in this paper has been quite general and should be straightforward to
generalize to small field inflationary models. Whether the positive features of
large field models carry over to small field ones remains an open question
requiring further examination.
Non-canonical scalar fields violate the consistency relation $r = -8n_T$, which emerges as a {\em smoking gun} test
for this class of models. Another test could be the extent of
 non-Gaussianity in the perturbation spectrum which we shall revert to in a future work.

After this paper was completed we became aware of \cite{liddle12} containing
results which partially overlapped with ours.

\acknowledgments
We acknowledge useful discussions with Charles Jose, Minu Joy, Jayanti Prasad, L.~Sriramkumar
and Tarun Souradeep.
We also benefitted from comments made by the referee
which led to an improvement in the manuscript.

\appendix

\section{Field redefinitions}

Under the field redefinition
\beq
\phi \rightarrow \psi \equiv b\,\phi,
\label{field redf-1}
\eeq
where $b$ is a constant, the Lagrangian (\ref{eqn: Lagrangian}) with a power law potential $V(\phi) = V_{_{0}}\,\phi^{n}$ becomes
\beq
{\cal L}(\widetilde{X},\psi) = \widetilde{X}\l(\frac{\widetilde{X}}{\widetilde{M}^{4}}\r)^{\alpha-1} -\; \tilde{V}_{_0}\,\psi^{n},
\eeq
where
$$\widetilde{X} \equiv \frac{1}{2}\pa_{\mu}\psi\; \pa^{\mu}\psi,$$
$$\widetilde{M} = M\,b^{(\alpha/2)/(\alpha -1)},$$
and
$$\tilde{V}_{_0}= \frac{V_{_0}}{b^{n}}.$$
We therefore find that the structure of the Lagrangian (2.13) with a power law potential is invariant under the  field redefinition $\phi \rightarrow b\,\phi$.
Only the parameters $M$ and $V_{_{0}}$  are influenced by
such a redefinition.
It is evident from Eqs.(\ref{eqn: ns phi n potential}) and (\ref{eqn: T-to-S phi-n pot}) that the values
 of the scalar spectral index $n_{_{S}}$ and the tensor to scalar ratio $r$ are independent of both $M$ and $V_{_{0}}$.
This implies that such observables  are immune to
the field redefinition $\phi \rightarrow b\,\phi$
 in the chaotic inflationary model based on (2.13).
(This statement might even be true for generic field redefinitions.)

Field redefinitions of the form (\ref{field redf-1}) do not affect
 the form of the potential.
However, with suitable transformations such as
$\phi \rightarrow \psi \equiv f(\phi)$, it is
 possible to convert a flat potential into a `steep' one
 {\em but with a different kinetic term} in the Lagrangian.
We illustrate this for the non-canonical scalar
field model (\ref{eqn: Lagrangian}) with a quadratic potential
\beq
{\cal L}(X,\phi) = X\l(\frac{X}{M^{4}}\r)^{\alpha-1}\; -\; \frac{m^{2}\phi^{2}}{2}\; .
\label{eqn: Lagrangian-A1}
\eeq
It is straightforward to show that under the transformation
\beq
\phi \rightarrow \psi \equiv \phi^{-1/s} \hspace{1cm} \mathrm{where} \hspace{1cm} s > 0,
\eeq
the Lagrangian (\ref{eqn: Lagrangian-A1}) becomes
\beq
{\cal L}(\widetilde{X},\psi) = f(\psi)\widetilde{X}\l(\frac{\widetilde{X}}{M^{4}}\r)^{\alpha-1} -\; \frac{V_{_0}}{\psi^{2s}},
\label{eqn: Lagrangian-A2}
\eeq
where
\beq
f(\psi) = \l(\frac{s}{\psi^{s + 1}}\r)^{2\alpha}~.
\eeq
Therefore, although the potential in (\ref{eqn: Lagrangian-A2}) is steep for $s > 0$, the structure of the kinetic term is different from that in (\ref{eqn: Lagrangian}).
It is therefore important to note that the steep potentials discussed in
 this paper are considered in the context of a given
 form of the kinetic term, namely the one described by (\ref{eqn: Lagrangian}).

\end{document}